%% file: ICDE_main.tex
\documentclass[conference]{IEEEtran}
\IEEEoverridecommandlockouts

\usepackage{cite}
\usepackage{marvosym}
\usepackage{amsmath}
\usepackage{amsmath,amssymb,amsfonts}
\usepackage{algorithmic}
\usepackage{graphicx}
\usepackage{textcomp}
\usepackage{xcolor}
\usepackage{caption}
\usepackage{subcaption}
\usepackage[linesnumbered,ruled,noend]{algorithm2e}
\usepackage{setspace}
\usepackage{enumitem}
\usepackage{stfloats}
\usepackage{pifont}
\usepackage{url}
\usepackage{balance}
\usepackage[normalem]{ulem}
\usepackage{multicol}
\usepackage{multirow}
\usepackage{booktabs}
\usepackage{makecell}
\usepackage{utfsym}
\usepackage{xspace}
\usepackage{framed}
\usepackage{breqn}
\usepackage{tabularx}
\usepackage{tcolorbox}
\tcbuselibrary{breakable}
\usepackage{amsthm}
\usepackage{ifthen}
\usepackage{hyperref}

\usepackage{balance}

\usepackage{marginnote}

\hypersetup{
    pdftitle={Efficient Query Rewrite Rule Discovery via Standardized Enumeration and Learning-to-Rank(extend)},
    pdfauthor={Yuan Zhang et al.},
    pdfsubject={ICDE Conference Paper},
    pdfcreator={LaTeX with hyperref}
}

\newtheorem{definition}{Definition}
\newtheorem{example}{Example}

\newcommand\sysname{\textsc{SLER}\xspace}

\newif\ifextended\extendedfalse

\def\BibTeX{{\rm B\kern-.05em{\sc i\kern-.025em b}\kern-.08em
    T\kern-.1667em\lower.7ex\hbox{E}\kern-.125emX}}

\begin{document}

\title{Efficient Query Rewrite Rule Discovery via Standardized Enumeration and Learning-to-Rank(extend)}

\author{
\IEEEauthorblockN{
Yuan Zhang$^{*}$,
Yuxing Chen$^{\dagger}$ \textsuperscript{\Letter},
Yuekun Yu$^{*}$,
Jinbin Huang$^{*}$,
Rui Mao$^{*}$ \textsuperscript{\Letter},\\
Anqun Pan$^{\dagger}$,
Lixiong Zheng$^{\dagger}$,
Jianbin Qin$^{*} \textsuperscript{\Letter}$
}
\IEEEauthorblockA{
$^{*}$Shenzhen Institute of Computing Science, Shenzhen University, Shenzhen, China \\
Email: \{2300271049,2410104038\}@email.szu.edu.cn,  \{jbhuang,mao,qinjianbin\}@szu.edu.cn
}
\IEEEauthorblockA{
$^{\dagger}$Tencent Inc., Shenzhen, China \\
Email: \{axingguchen,aaronpan,paterzheng\}@tencent.com
}
}


\maketitle
\thispagestyle{plain}
\pagestyle{plain}

\begin{abstract}

Query rewriting is essential for database performance optimization, but existing automated rule enumeration methods suffer from exponential search spaces, severe redundancy, and poor scalability, especially when handling complex query plans with five or more nodes, where a \emph{node} represents an operator in the plan tree. We present \sysname, a scalable system that enables efficient and effective rewrite rule discovery by combining standardized template enumeration with a learning-to-rank approach. \sysname uses \emph{standardized templates}, abstractions of query plans with operator structures preserved but data-specific details removed, to eliminate structural redundancies and drastically reduce the search space. A learn-to-rank model guides enumeration by pre-filtering the most promising template pairs, enabling scalable rule generation for large-node templates.

Evaluated on over 11,000 real-world SQL queries from both open-source and commercial workloads, \sysname has automatically constructed a rewrite rule repository exceeding 1 million rules—the largest empirically validated rewrite rule library to date. Notably, at the scale of one million rules, \sysname supports query plan templates with complexity up to \emph{channel}-level depth. This unprecedented scale opens the door to discovering highly intricate transformations across diverse query patterns. Critically, \sysname's template-driven design and learned ranking mechanism are inherently extensible, allowing seamless integration of new and complex operators, paving the way for next-generation optimizers powered by comprehensive, adaptive rule spaces.

\end{abstract}


\section{INTRODUCTION}\label{sec:introduction}
\input{INTRODUCTION}

\section{RELATED WORK}\label{sec:related_work}
\input{Relatedwork}

\section{PRELIMINARY AND OVERVIEW}\label{Char : Rule Mining Framework}
\input{PRELIMINARY-AND-OVERVIEW}

\section{STANDARDIZATION RULES}\label{Char: OPTIMISATION}
\input{Optimisation-based-on-standardisation}

\section{RULE RANKER}\label{Char: SIMPLICITY OF RULES}
\input{Simplicity-of-rules}

\section{EVALUATION}\label{Char: Evaluation}
\input{EVALUATION}

\section{Discussion}\label{sec:discussion}







\noindent \textbf{Scope and Limitations.} While \sysname shows strong performance in scaling rule discovery, its current implementation has specific scopes. (1) \textit{Operator Support}: We currently support core relational operators, including Input, Project, Filter, (Inner/Left/Right) Join, InSub, and Distinct. These form the backbone of most SPJ-style (Select-Project-Join) queries. (2) \textit{Beneficiary Patterns}: Our framework is particularly effective for ORM-generated SQL queries, which often feature deeply nested subqueries and large-scale structural redundancies that exceed the optimization window of traditional 4-node rule sets. (3) \textit{Extensibility}: To extend \sysname to aggregates, group-by, and window functions, two main requirements must be met: first, defining the First-Order Logic (FOL) semantics for these operators to enable Z3 verification; second, establishing the corresponding standardized templates for enumeration. This represents a promising direction for future work to broaden \sysname's applicability to complex OLAP workloads.
\noindent \textbf{The Practical Bound of Enumeration.} While the enumeration space grows exponentially with the number of nodes, we observe that nodes between 7 and 9 represent a ``spot'' for rule discovery. Beyond 10 nodes, rules become overly specific and rarely triggered in real-world queries. By focusing on these complex yet reusable patterns, SLER avoids the infinite pursuit of extremely large templates while significantly outperforming the 4-node limit of SOTA works. Most importantly, the enumeration is a one-time process. Once the rules are discovered, they can be directly integrated into DBMSs.


\section{Conclusion}
\input{Conclusion}

\section*{AI-Generated Content Acknowledgement}
We used Gemini to edit the main text. AI was only used for polishing the writing. We are responsible for all the materials presented in this work.
\balance

\bibliographystyle{IEEEtran}
\bibliography{Reference}

\end{document}

%% file: INTRODUCTION.tex
\begin{sloppypar}

Query rewriting is a critical technique for enhancing the performance of database-driven applications~\cite{tang2023detecting,bai2023querybooster,selinger1979access}, as it optimizes poorly constructed queries before they are processed by the query optimizer~\cite{graefe1987rule,graefe1993volcano,graefe1995cascades}. Traditional approaches rely on manually derived rewrite rules, which are time-consuming to develop and unsustainable for evolving workloads~\cite{graefe1995cascades,graefe1987exodus,graefe1993volcano,levy1997query,mumick1990magic,muralikrishna1992improved,pirahesh1992extensible,seshadri1996cost}. 

The state-of-the-art (SOTA) rule enumeration engine, WeTune \cite{wang2022wetune}, generates query rewrite rules by automatically enumerating pairs of query plan templates (typically modeled as plan trees with operators as nodes) and their associated constraints. However, it employs a brute-force approach to enumerate all possible rules, resulting in prohibitively high computational complexity. For rules involving four or fewer nodes, it requires 4320 CPU hours, equivalent to approximately nine days. 
Worse, over 90\% of the rules it mines are redundant, and over 50\% of the remaining non-redundant rules are trivial, offering minimal optimization value for database system performance. Scaling to more complex rules is also infeasible, as enumerating five-node rules with WeTune would take approximately six months on a 64-CPU machine, and six-node rules could require a decade or more.

This inefficiency is particularly problematic given the characteristics of real-world queries. Analyses of web application and benchmark workloads \cite{tpch2023,tpcds2024} reveal that most queries involve more than four nodes. Production databases like PostgreSQL\cite{postgresql} frequently optimize plans containing 5 to 8 nodes, while complex queries may exceed 10 nodes. Thus, WeTune’s rule base, limited to 4- or fewer-node patterns, is insufficient for optimizing complex queries, rendering it inadequate for practical scenarios \cite{li2024llm}.
\begin{table*}[!htbp]
\centering
\scriptsize 
\setlength{\tabcolsep}{3pt} 
\caption{Example queries and their rewritten forms by WeTune~\cite{wang2022wetune} versus our method (\sysname). WeTune fails to rewrite $q_1$, transforms $q_7$ into a non-minimal form, and incurs higher rewriting overhead for $q_5$ despite equivalence to \sysname's output.}
\label{tab: example of query}
\resizebox{\textwidth}{!}{ 
\begin{tabular}{|p{0.32\textwidth}|p{0.32\textwidth}|p{0.32\textwidth}|} 
\toprule 
\textbf{Original Query} & \textbf{WeTune} & \textbf{Ideal (\sysname)} \\
\midrule 
\makecell[l]{
$q_1$: \textcolor{red}{SELECT} e1.dept\_id \textcolor{red}{AS} s2 \textcolor{red}{FROM} (\\
\ \ \textcolor{red}{SELECT} dept\_id \textcolor{red}{AS} s0, region \\
\ \  \textcolor{red}{FROM} Employees\\
\ \ \textcolor{red}{WHERE} salary > 50000) e1\\
\ \ \textcolor{red}{LEFT JOIN} (\\
\ \ \ \ \textcolor{red}{SELECT} region \textcolor{red}{AS} s1, dept\_id \\
\ \ \ \  \textcolor{red}{FROM} Employees) e2\\
\ \ \textcolor{red}{ON} e1.region = e2.region
} &
\makecell[l]{
$q_2$: \textcolor{red}{SELECT} e1.dept\_id \textcolor{red}{AS} s2 \textcolor{red}{FROM} (\\
\ \ \textcolor{red}{SELECT} dept\_id \textcolor{red}{AS} s0, region \\
\ \ \textcolor{red}{FROM} Employees\\
\ \ \textcolor{red}{WHERE} salary > 50000) e1\\
\ \ \textcolor{red}{LEFT JOIN} (\\
\ \ \ \ \textcolor{red}{SELECT} region \textcolor{red}{AS} s1, dept\_id \\
\ \ \ \ \textcolor{red}{FROM} Employees) e2\\
\ \ \textcolor{red}{ON} e1.region = e2.region
} &
\makecell[l]{
$q_3$: \textcolor{red}{SELECT} dept\_id \textcolor{red}{AS} s2\\
\ \ \textcolor{red}{FROM} Employees\\
\ \ \textcolor{red}{WHERE} salary > 50000
} \\
\hline
\makecell[l]{
$q_4$: \textcolor{red}{SELECT DISTINCT} s.Region, s.TotalSales \\
\ \ \textcolor{red}{FROM} Sales s\\
\ \ \textcolor{red}{WHERE} s.Category = '\textcolor{blue}{Electronics}' \\
\ \ \textcolor{red}{AND} s.Region \textcolor{red}{IN} (\\
\ \ \ \ \textcolor{red}{SELECT} n.Region \textcolor{red}{FROM} Sales n\\
\ \ \ \ \textcolor{red}{WHERE} n.Category = '\textcolor{blue}{Electronics}')
} &
\makecell[l]{
$q_5$: \textcolor{red}{SELECT DISTINCT} s.Region, s.TotalSales\\
\ \ \textcolor{red}{FROM} Sales s\\
\ \ \textcolor{red}{WHERE} Category = '\textcolor{blue}{Electronics}'
} &
\makecell[l]{
$q_6$: \textcolor{red}{SELECT DISTINCT} s.Region, s.TotalSales\\
\ \ \textcolor{red}{FROM} Sales s\\
\ \ \textcolor{red}{WHERE} Category = '\textcolor{blue}{Electronics}'
} \\
\hline
\makecell[l]{
$q_7$: \textcolor{red}{SELECT} e1.emp\_id, e1.name, e1.dept\_id \\
\ \ \textcolor{red}{FROM} Employees e1\\
\ \ \textcolor{red}{INNER JOIN} (\\
\ \ \ \ \textcolor{red}{SELECT DISTINCT} dept\_id \textcolor{red}{AS} s0 \\
\ \ \textcolor{red}{FROM} Projects) p\\
\ \ \textcolor{red}{ON} e1.dept\_id = p.dept\_id\\
\ \ \textcolor{red}{WHERE} e1.dept\_id \textcolor{red}{IN} (\\
\ \ \ \ \textcolor{red}{SELECT} e2.dept\_id \textcolor{red}{AS} s1 \\
\ \ \ \ \textcolor{red}{FROM} Employees e2\\
\ \ \ \ \textcolor{red}{LEFT JOIN} Projects p2\\
\ \ \ \ \textcolor{red}{ON} e2.dept\_id = p2.dept\_id)
} &
\makecell[l]{
$q_8$: \textcolor{red}{SELECT} e1.emp\_id, e1.name, e1.dept\_id\\
\ \ \textcolor{red}{FROM} Employees e1\\
\ \ \textcolor{red}{INNER JOIN} (\\
\ \ \ \ \textcolor{red}{SELECT DISTINCT} e2.dept\_id \textcolor{red}{AS} s0\\
\ \ \ \ \textcolor{red}{FROM} Projects p \\
\ \ \ \ \textcolor{red}{INNER JOIN} Employees e2\\
\ \ \ \ \textcolor{red}{ON} p.dept\_id = e2.dept\_id) p\\
\ \ \textcolor{red}{ON} e1.dept\_id = p.dept\_id
} &
\makecell[l]{
$q_9$: \textcolor{red}{SELECT} e.emp\_id, e.name, e.dept\_id, \\
\ \ p.dept\_id  \textcolor{red}{AS} s0\\
\ \ \textcolor{red}{FROM} Employees e\\
\ \ \textcolor{red}{LEFT JOIN} (\\
\ \ \ \ \textcolor{red}{SELECT DISTINCT} dept\_id \textcolor{red}{AS} s0\\
\ \ \ \ \textcolor{red}{FROM} Projects) p\\
\ \ \textcolor{red}{ON} e.dept\_id = p.dept\_id
} \\
\bottomrule 
\end{tabular}
} 
\end{table*}

\vspace{2mm}
\noindent \textbf{Challenges.}  Existing SOTA works like WeTune on query rewrite rule exploration exhibit limitations in both efficiency and effectiveness due to two key deficits: (1) a lack of effective rules for specific complex patterns, resulting in failures to simplify intricate queries \textbf{(ineffectiveness)}; (2) the need for multiple rule-identifying iterations to optimize queries with more than four nodes \textbf{(inefficiency)}, coupled with its constrained enumeration space, which prevents reduction to the simplest form. Examples of these limitations are provided below.

\begin{example}
    Table \ref{tab: example of query} showcases three original queries (more detailed description in \S \ref{sec:related_work}), alongside their rewritten versions generated by WeTune and their corresponding ideal forms. The ideal form, defined as the syntactically simplest representation, typically exhibits superior performance in database systems relative to the original query structure. WeTune fails to rewrite query \(q_1\) due to the absence of applicable transformation rules in its rule base. Although WeTune’s rewritten form of query \(q_5\) aligns with the ideal solution, the computational overhead incurred during rewriting is notably higher. For query \(q_7\), WeTune can only produce a non-minimal form. This stems from its constrained rule base, which is primarily designed for query patterns with four or fewer nodes. 
\end{example}

Additionally, assessing rule effectiveness using formal methods like SMT solvers (e.g., Z3~\cite{deMoura_Bjrner_2008}) is constrained by unresolved complexity questions such as NP=coNP~\cite{pudlak1998lengths,krajivcek2019proof,cook1979relative}, making proof lengths unreliable metrics for rule quality, especially when query plans are with larger node sizes. Enumerating rules for templates with >6 nodes faces combinatorial explosion, as 99\% of time (couple of years \cite{wang2022wetune}) is consumed by verification calls, rendering existing methods impractical.

\vspace{2mm}
\noindent \textbf{Contributions.} This paper proposes SLER, a query rewrite framework consisting of a rule enumerator, deduplicator, and ranker, designed to efficiently identify effective rules that can improve database system performance, addressing the key challenges of efficiency and effectiveness in query rewrite rule identification.

\vspace{1mm}
(1) \textbf{Efficiency.}  Compared to the brute-force strategy in WeTune, the ``small-to-large'' hypothesis (details in \S \ref{sec:preliminary_motivatin}), which suggests deriving complex rules by composing simpler ones, has been proven ineffective. Empirical analysis shows that only 9\% of 4$\to$2 node rules can be composed from 4$\to$3 and 3$\to$2 node rules, confirming that complex rules require direct enumeration. To address this, our \textit{enumerator} (\S \ref{sec:Standardisation of rule templates})  employs standardized plan templates for rule generation. In addition, we define a \textit{ranker} (\S \ref{Char: SIMPLICITY OF RULES}) that adopts a learning-to-rank model~\cite{burges2010ranknet,chapelle2011empirical} to prioritize plan templates based on their effectiveness in optimizing query performance.
The enumerator first enumerates all plan templates, then selectively applies the ranker filtering mechanism to select the top-$k$ plan template pairs. A larger $k$ increases the proportion of effective rules in the final set but may increase enumeration time. This pre-filtering discards ineffective or redundant templates early, avoiding unnecessary query equivalence verification (which typically requires exponential Z3 assertions~\cite{deMoura_Bjrner_2008}). Then, source and destination templates are instantiated as queries, validated against standardized rules, and normalized to ensure consistency. To ensure the generation of all effective rules, enumeration can also be performed without using the ranker.
To further optimize, our \textit{deduplicator} (\S \ref{sec:Rule De-Redundancy}) implements the Reduced by Template Pair (RTP) algorithm, pruning redundant rules through regularization and deduplication. The deduplicator initializes an empty rule base, verifies the validity of generated rules, and checks for redundancy among checksum-equivalent rules, thereby significantly reducing the candidate set while maintaining optimization capability.

\vspace{1mm}
(2) \textbf{Effectiveness.} SMT-based proof systems (e.g., Z3) struggle to reliably measure rule complexity due to unresolved theoretical issues such as NP=coNP, hindering effective rule prioritization.
Our \textit{ranker} (\S \ref{Char: SIMPLICITY OF RULES}) focuses on validating rule sets by screening for potentially effective rules and avoids validating ineffective ones through features such as operator types and corresponding expression complexities. The model is trained on over 11,000 real-world SQL queries from a production database and can predict rule scores. During inference, rules are sorted in descending order of scores to highlight those with the greatest performance impact. Unlike brute-force methods limited to small-scale ($\leq$4 node) queries, this ranker efficiently handles both small-node and large-node queries (previously inaccessible), resolving scalability issues and expanding analyzable query patterns through integrated rule validity screening and deduplication techniques.

We summarize our contributions as follows:
\begin{itemize}[leftmargin=*]
\item Formalize the ``small-to-large'' hypothesis and propose a standardized rule enumeration algorithm to reduce enumeration overhead and improve node coverage while balancing rule effectiveness (\S \ref{sec:Standardisation of rule templates})..
\item Develop new rule-based redundancy elimination optimization algorithms, incorporating normalization, regularization, and structural similarity for efficient processing and candidate set reduction (\S \ref{sec:Rule De-Redundancy}).
\item Reveal the limitations of mathematical methods for rule prioritization and introduce a machine learning-based ranking mechanism with plan template validity screening to enhance rule effectiveness in database query optimization and selectively reduce overhead (\S \ref{Char: SIMPLICITY OF RULES}).
\item Evaluate SLER on open-source and commercial queries, demonstrating its efficiency and effectiveness superiority over the SOTA WeTune (\S \ref{Char: Evaluation}).
\end{itemize}

\end{sloppypar}

%% file: Relatedwork.tex
\textbf{Query optimization} remains a cornerstone challenge in database systems, with research spanning three interconnected domains: cardinality estimation (e.g., \cite{kipf2018learned,sun2021learned,sun2021learned2,wu2018towards,yang2019deep}), cost estimation (e.g., \cite{marcus2019plan,sun2019end,liu2015forecasting,wu2013predicting}), and query plan search (e.g., \cite{trummer2021skinnerdb,yu2020reinforcement,stillger2001leo,krishnan2018learning,marcus2019neo,yang2022balsa,marcus2021bao,lohman1988grammar,seshadri1996cost}). Collectively, these areas strive to enhance query efficiency by refining input accuracy, evaluating plan efficacy, and exploring optimal execution strategies~\cite{sun2021learned,sun2021learned2,dintyala2020sqlcheck}, yet critical gaps persist in handling complex, real-world workloads~\cite{marcus2019plan,sun2019end}. Recent research has explored query rewriting rules using large language models (LLMs), addressing the limitations of traditional methods in processing diverse and complex patterns. LLMs leverage common knowledge to achieve improved generalization~\cite{liu2024query,sun2024r,li2024llm,dharwada2025query}. These methods demonstrate the potential of LLMs in identifying redundancy and optimizing performance.
However, these studies are mostly based on improvements to existing rules or evidence bases, making it difficult to automatically discover new rules, often subject to hallucinations, and lacking stability.~\cite{bai2023demo} which allows for custom rules, however, requires a high level of expertise and deep domain knowledge, resulting in a high barrier to entry and limiting its widespread adoption among non-expert users. Notably, these methods are orthogonal to the query rewrite explored in this paper and can be applied subsequent to our proposed approach.

\vspace{2mm}
\noindent \textbf{Query rewriting} constitutes a pivotal technique for improving the performance of database-driven applications~\cite{tang2023detecting}, as it optimizes poorly constructed queries before they are processed by the query optimizer~\cite{graefe1987rule,graefe1993volcano,graefe1995cascades}. Traditional approaches rely on manually derived rewrite rules, which are time-intensive to develop and unsustainable for evolving workloads. Despite decades of development, existing production-ready database systems (e.g., Oracle \cite{oracle_db}, DB2 \cite{ibm_db2}, TDSQL \cite{DBLP:journals/pvldb/ChenPLYHTLCZD24_TDSQL,DBLP:conf/sigmod/WangCJPJWLZZLCZ25_TXSQL}, SQL Server \cite{microsoft_sql_server}, Amazon RDS \cite{amazon_rds}) rely on extensive query rewrite rule sets, yet many queries remain unoptimized. This observation suggests that a significant number of effective rewrite rules may still be undiscovered, highlighting a potential gap in current query optimization strategies. 

Research by \cite{negi2021steering} suggests that a tailored subset of rewrite rules can outperform a comprehensive set for targeted workloads, improving plan quality while reducing optimizer overhead. \cite{zhou2021learned} highlights the significance of the order in which rewrite rules are applied. However, both approaches are limited in their capacity to expand the query plan search space or propose new rules, thereby limiting their effectiveness.
Current SOTA work, i.e., WeTune \cite{wang2022wetune}, has made progress in automative rule generation but faces significant limitations, including expansive enumeration spaces, inefficient processing, and pervasive redundant rules. To address the above, this paper aims to improve the efficiency and effectiveness of query rewrite rule enumeration. 

\begin{sloppypar}

\vspace{3mm}
\noindent \textbf{Research Gaps.} 
Firstly, traditional reliance on manual derivation of query rewrite rules, driven primarily by programmer intuition, suffers from inherent incompleteness. This approach fails to exhaustively identify all potential rules, leaving a significant portion of optimization opportunities unexplored. The gap is exacerbated by the proliferation of auto-generated SQL: queries produced by AI tools \cite{openai_chatgpt} and object-relational mapping (ORM) frameworks often exhibit counterintuitive structures (e.g., redundant joins or unnecessary subqueries). These auto-generated queries remain poorly optimized by existing database systems, as the manually derived rules in current optimizers are not designed to address their unique patterns. 
For instance, Table \ref{tab: example of query} shows three queries that retrieve department IDs for employees with salaries above 50,000 using redundant subqueries and a self-join on regions (\(q_1\)), fetch unique regions and total sales for electronics with a redundant subquery (\(q_4\)), and select employee details for departments in projects with unnecessary subqueries and joins (\(q_7\)), respectively, and are common outputs of ORM tools. 
The nested $q_1$ cannot be simplified by standard rewrite rules, despite its obvious inefficiency. 

Secondly, SOTA techniques like WeTune \cite{wang2022wetune}, are constrained by two key limitations. First, they are restricted to generating rules for queries with up to four nodes, yet real-world databases (e.g., PostgreSQL \cite{postgresql}) frequently process queries with 5–8 nodes (operators), and complex workloads may involve 10+ nodes. This mismatch renders existing methods inadequate for optimizing large-scale queries. Second, WeTune’s exhaustive search approach is computationally inefficient. WeTune requires over 9 days to generate rules for 4- or few-node queries, with 90\% redundancy (redundant example rules in Table \ref{tab: example of redundant rules}). Our attempts to mitigate this via a ``small-to-large'' strategy, deriving complex rules from simpler ones, have proven ineffective (\S \ref{sec:preliminary_motivatin}). These limitations underscore the need for a more scalable method to enumerate complex rules.

Thirdly, existing methods struggle to distinguish impactful rules from trivial ones that offer minimal optimization benefits. Mathematical approaches, such as SMT solvers like Z3~\cite{deMoura_Bjrner_2008}, fail to reliably evaluate rule effectiveness due to unpredictable proof lengths caused by the NP=coNP problem~\cite{pudlak1998lengths,krajivcek2019proof,cook1979relative}. For example, rules that only rename attributes without altering logical structure 
are often included in rule sets despite providing no meaningful optimization. This lack of robust filtering mechanisms clogs optimizers with redundant rules, increasing overhead and reducing efficiency. Thus, a critical gap exists in developing scalable techniques to prioritize rules based on their actual optimization value.

\end{sloppypar}

%% file: PRELIMINARY-AND-OVERVIEW.tex
\begin{sloppypar}

This section introduces key concepts and formal definitions used throughout the paper. It is followed by an introduction to the motivation and overall design of this paper.

\subsection{Preliminary}
A query plan template ($T$) is defined as a directed tree, where each node corresponds to a relational algebra operator.  Except for the Input operator, each operator takes one or two relations as input, executes algebraic computations via its semantics, and outputs a single relation. In line with WeTune \cite{wang2022wetune}, the operators supported in this work are as follows:


\begin{itemize}[leftmargin=*]
    \item Input(\( R \)): Represents an input relation, parameterized by a symbolic relation \( R \) (e.g., a table or subquery result).
    \item Project(\( a \)): Projects tuples to retain only the attributes specified in the list \( a \) (e.g., selecting specific columns from a table).
    \item Filter(\( p, a \)): Filters out tuples that do not satisfy predicate \( p \), where $p$  is evaluated using attributes in the list \( a \) (e.g., retaining rows where ``age > 30''). 
    \item InnerJoin(\( R_l.a_l, R_r.a_r \)): Performs an inner join between the relations $R_l$ (left) and $R_r$ (right), retaining tuples where attributes $R_l.a_l$ and $R_r.a_r$ match.
    \item LeftJoin$(R_l.a_l,R_r.a_r)$: Performs a left join between the relations $R_l$ and $R_r$, retaining tuples where attributes  $R_l.a_l$ and $R_r.a_r$ match.
    \item RightJoin$(R_l.a_l,R_r.a_r)$: symmetric to LeftJoin
    \item InSub(\( R_l.a_l, R_r.a_r \)): Filters tuples from \( R_l \) to retain only those where the value in \( R_l.a_l \) exists in \( R_r.a_r \) (e.g., where id in (select id from ...)).
    \item Distinct(\( a \)): Projects attributes \( a \) and removes duplicate tuples (equivalent to SQL’s DISTINCT keyword).
\end{itemize}

Constraints in constraint set \( C \) ensure that \( T_{\text{src}} \) and \( T_{\text{dest}} \) are semantically equivalence. 
They define relationships $M$ between symbols (relations, attributes, predicates) in \( T_{\text{src}} \) and \( T_{\text{dest}} \):

\begin{itemize}[leftmargin=*]
    \item RelEq(\( R_1, R_2 \)): Relations \( R_1 \) and \( R_2 \) contain the identical tuples  (i.e., they are logically the same dataset).
    \item AttrsEq(\( a_1, a_2 \)): Attribute lists \( a_1 \) and \( a_2 \) are identical (e.g., both list \([id, name]\)).
    \item PredEq(\( p_1, p_2 \)): Predicates \( p_1 \) and \( p_2 \) are logically equivalent, i.e., \( p_1 \Leftrightarrow p_2 \) (e.g., \(p_1 = (x \ge 5)\) and \(p_2 = (x \geq 5)\) for integer x).
    \item SubAttrs(\( a_1, a_2 \)): Each attribute in \( a_1 \) is also in \( a_2 \) (e.g., \(a_1 = [id]\) and \(a_2 = [id, name]\)).
    \item RefAttrs(\( R_1, a_1, R_2, a_2 \)): All values in \( R_1 \) on \( a_1 \) are also in \( R_2 \) on \( a_2 \) (e.g., all user IDs in \(R_1\) exist in \(R_2\)).
    \item Unique(\( R, a \)): Values in \( R \) on \( a \) are unique (e.g., a primary key \( a \)).
    \item NotNull(\( R, a \)): No value in \( R \) on \( a \) is NULL (e.g., a column with a ``NOT NULL'' constraint).
\end{itemize}

A \textbf{query rewrite rule} formalizes how a source template can be transformed into a semantically equivalent destination template under specific constraints:
\begin{definition}[Query Rewrite Rule]

A query rewrite rule is $R = (T_{\text{src}}, T_{\text{dest}}, M, C)$, where:

\begin{itemize}[leftmargin=*]

    \item $T_{\text{src}}, T_{\text{dest}}$ are the source and destination templates, respectively.
    
    \item $M: V_{\text{src}} \to V_{\text{dest}}$ is a mapping associating nodes in $T_{\text{src}}$ to $T_{\text{dest}}$.
    
    \item $C$ is a set of constraints that guarantee \(T_{\text{src}}\) and \(T_{\text{dest}}\) are semantically equivalent.
    
\end{itemize}

\end{definition}

\begin{figure}[!t]
  \centering
  \includegraphics{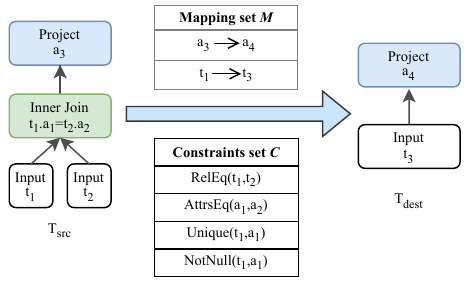}
  \caption{Example of a rule model.}\label{fig:rule_example}
\end{figure}

A rule of $R$ is valid if the following two conditions are met: 
(1)  \(T_{\text{dest}}\) contains fewer nodes (operators) than that \(T_{\text{src}}\);
(2) $C$ is minimal in enabling the equivalence between \(T_{\text{src}}\) and \(T_{\text{dest}}\). Formally, $C$ must be the most relaxed set of constraints such that removing any single constraint from $C$ would break the semantic equivalence of the \(T_{\text{src}}\) and \(T_{\text{dest}}\).

If both conditions hold, \(T_{\text{src}}\) is replaced with \(T_{\text{dest}}\) to form an optimized query. 
The mapping \( M \) formalizes the transformation applied by rule \( R \). 
For instance, Figure~\ref{fig:rule_example} illustrates a sample rule where a join operation in \(T_{\text{src}}\) is eliminated in \(T_{\text{dest}}\) under specific constraints (e.g., uniqueness of a join attribute). In this case, the mapping \( M \) means that the relation \( t_3 \) in \( T_{\text{src}} \) is substituted with \( t_1 \), attribute \( a_4 \) is replaced by \( a_3 \), and \( M \) is omitted from \( R \) sometimes, to improve readability. In summary, the primary notations are shown in Table \ref{tab:notations}.

\begin{table*}[!t]
\centering
\caption{Example of redundant rules: WeTune generates all three, with \(R_1\) and \(R_2\) redundant, whereas \sysname retains only \(R_3\).}
\label{tab: example of redundant rules}
\footnotesize
\setlength{\tabcolsep}{4pt} 
\renewcommand{\arraystretch}{1.15} 

\resizebox{0.95\textwidth}{!}{%
\begin{tabular}{|c|c|c|c|}
\hline
\multicolumn{1}{|c|}{\textbf{Rule}} &
\multicolumn{1}{c|}{\textbf{Source Plan Template}} &
\multicolumn{1}{c|}{\textbf{Destination Plan Template}} &
\multicolumn{1}{c|}{\textbf{Extra Constraints}} \\
\hline
$R_1$ &
$\mathrm{Proj_{t_0.a_0}(IJoin_{t_0.a_1,t_0.a_0}(Proj_{t_0.a_0}(t_0),Dedup(Proj_{t_0.a_1}(t_0))))}$ &
$\mathrm{Proj_{t_0.a_0}(t_0)}$ &
$\mathrm{NotNull(t_0,a_0);Unique(t_0,a_1)}$ \\
\hline
$R_2$ &
$\mathrm{Proj_{t_0.a_1}(IJoin_{t_0.a_0,t_0.a_1}(Proj_{t_0.a_0}(t_0),Dedup(Proj_{t_0.a_1}(t_0))))}$ &
$\mathrm{Proj_{t_0.a_0}(t_0)}$ &
$\mathrm{NotNull(t_0,a_1);Unique(t_0,a_1)}$ \\
\hline
$R_3$ &
$\mathrm{Proj_{t_0.a_0}(IJoin_{t_0.a_0,t_0.a_1}(Proj_{t_0.a_0}(t_0),Dedup(Proj_{t_0.a_1}(t_0))))}$ &
$\mathrm{Proj_{t_0.a_0}(t_0)}$ &
$\mathrm{NotNull(t_0,a_0);NotNull(t_0,a_1);Unique(t_0,a_1)}$ \\
\hline
\end{tabular}%
}
\end{table*}

\begin{table}[!t]
  \caption{Summary of Key Notations.}\label{tab:notations}
  \belowrulesep=0pt
  \aboverulesep=0pt
  \centering
  \footnotesize 
  \renewcommand{\arraystretch}{1.2}
  
  \begin{tabular}{cc} 
    \toprule
    \textbf{Notation} & \textbf{Description} \\ \midrule
    $n$ & Node count (template size) \\ \midrule
    $T_{src}, T_{dest}$ & Source and destination templates \\ \midrule
    $C$ & Validity constraints (schema, predicates) \\ \midrule
    $M$ & Structural mappings of operators \\ \midrule
    $r$ & Verified rule tuple: $(T_{src}, T_{dest}, C, M)$ \\ \midrule
    $k$ & Top-$k$ filtering threshold in LTR \\ \bottomrule
  \end{tabular}
  \vspace{-4mm}
\end{table}

\subsection{Motivation}\label{sec:preliminary_motivatin}
To derive a not only valid but also effective rule, the current methods discover potential rewrite rules through the following processes. 
First, all possible plan templates are enumerated. To keep the search space manageable, the size of these templates is constrained, i.e., the number of operators in each template is limited to a small threshold.
Next, all possible constraints are generated for each pair of enumerated plan templates. Finally, the most promising rules, that are the highest potential to improve query performance, are selected.
To reduce the computational cost of enumerating complex rules, an intuitive approach is to explore a compositional method that infers from larger nodes to smaller ones step by step, which we formalize as the small-to-large hypothesis.










\begin{definition}[Small-to-Large Composition]

Let $\mathcal{R}_{n \to m}$ denote the set of query rewrite rules transforming an $n$-node plan to an $m$-node plan ($m < n$). For $m < k < n$, a composition operator $\circ: \mathcal{R}_{n \to k} \times \mathcal{R}_{k \to m} \to \mathcal{R}_{n \to m}$ is defined such that $R_{n \to m} = R_{n \to k} \circ R_{k \to m}$ if applying $R_{n \to k}$ followed by $R_{k \to m}$ yields $R_{n \to m}$. The small-to-large hypothesis holds if $|\mathcal{R}_{n \to m}| = |\mathcal{R}_{k \to m}|$ and $\mathcal{R}_{n \to m} \subseteq \{ R_{n \to k} \circ R_{k \to m} \mid R_{n \to k} \in \mathcal{R}_{n \to k}, R_{k \to m} \in \mathcal{R}_{k \to m} \}$.

\end{definition}

Empirically, this hypothesis often fails. For example, only 9\% of rules in $\mathcal{R}_{4 \to 2}$ can be composed from $\mathcal{R}_{4 \to 3}$ and $\mathcal{R}_{3 \to 2}$. This indicates most complex rules require direct enumeration rather than composition. WeTune addresses this with brute-force enumeration but struggles to verify rules with five or more nodes. The root cause lies in WeTune’s inevitable generation of redundant rules. As shown in Table \ref{tab: example of redundant rules}, Rules \(R_1\) and \(R_2\) are redundant relative to Rule \(R_3\).
Despite differences in join conditions or projected attributes in their source templates, they produce semantically equivalent results under unique and non-null constraints. Yet WeTune generates all three rules exhaustively while our proposed standardized enumeration selectively retains only \(R_3\) (later in \S \ref{sec:Standardisation of rule templates}).

Enumerated rules are then typically verified for semantic equivalence using proof systems (e.g., SMT solvers like Z3 \cite{deMoura_Bjrner_2008}), during which the Z3 FOL is recorded as a metric for ``Verifying Complexity''.

\begin{definition}[Proof Length and Optimality]

For a set of query rewrite rules $\mathcal{R}$, let $V: \mathcal{R} \to \{\top, \bot\}$ be a verification function where $V(R) = \top$ if $R$ preserves query semantics. For a proof system $P$ (e.g., Z3~\cite{deMoura_Bjrner_2008}), $\ell_P(R)$ is the number of assertions required to prove $V(R) = \top$. $P$ is optimal if for any proof system $Q$, there exists a polynomial $p$ such that $\ell_P(R) \leq p(\ell_Q(R))$ for all $R \in \mathcal{R}$.

\end{definition}

Optimal proof systems are theoretical, as their realization hinges on unresolved complexity questions (e.g., NP=coNP \cite{pudlak1998lengths, krajivcek2019proof, cook1979relative}). 
WeTune generates all rules for verification, resulting in high verification overhead.
In contrast, \sysname can leverage rule pruning to identify optimal n-node rules (n ≤ 6), reducing verification time into polynomial time for redundant rules (\S \ref{sec:Rule De-Redundancy}), a significant improvement over the previous state-of-the-art exponential-time overhead. However, even \sysname's efficiency degrades significantly beyond six nodes, with verification times becoming impractical for n $\geq$ 7.
To address this, \sysname (\S ~\ref{Char: SIMPLICITY OF RULES}) uses a LambdaMART ranking model \cite{burges2010ranknet,chapelle2011empirical} to prioritize effective rules, leveraging features like tree distance and expression complexity. By leveraging the ranker to filter plan templates, we trade off the proportion of effective rules in the final set to reduce enumeration time for multi-node scenarios, thereby enabling the enumeration of rules with \( n \geq 7 \) nodes.  The LambdaMart is an ideal choice for \sysname rule ranking compared to traditional classification methods, regression methods \cite{burges2010ranknet}, deep learning methods \cite{marcus2021bao,marcus2018towards,he2014practical}, and other LTR methods \cite{chapelle2011yahoo,zhu2023lero,liu2009learning,cao2007learning} due to its optimization capabilities for ranking tasks, flexibility in handling complex features, and fit with query optimization requirements.


\subsection{System Overview}\label{sec: overview}
\begin{figure}[t]
  \centering
  \includegraphics{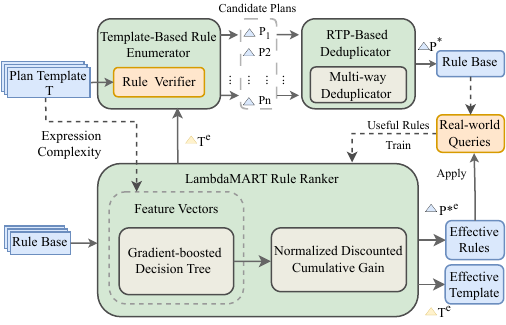}
  \caption{System architecture of \sysname.}\label{fig:overview}
\end{figure}
\sysname is designed to identify effective rules for improving the performance of real-world queries, as illustrated in Figure~\ref{fig:overview}.
Its workflow proceeds as follows: First, candidate standardized rules, explicitly designed to be non-redundant, are generated for each template, where these rules are fed into a rule equivalence checker to construct a standardized rule base. 
Subsequently, redundant rules are further Reduced by Template Pair (denoted as RTP algorithm). Lastly, the remaining rules, that are not possible to enumerate with higher nodes, are ranked by their effectiveness via a learning-to-rank model called LambdaMART ~\cite{burges2010ranknet,chapelle2011empirical}.During the enumeration process, for rules involving higher node counts, the ranking model can be applied to the plan templates first, followed by selecting the top-$k$ templates before proceeding with rule enumeration.


\textbf{Template-Based Rule Enumerator (\S \ref{sec:Standardisation of rule templates}).} 
To address the limitations of the brute-force enumeration, \sysname employs standardized templates for rule enumeration. Unlike WeTune enumerates all templates and yields exponential enumeration verifying complexity, \sysname can obtain polynomial-time verifying complexity for certain redundant templates.
The process begins by selecting two plan templates from the template library, designated as source and destination. 
Optionally, to reduce enumeration time and enable handling of larger node counts, the LambdaMART ranker (\S \ref{Char: SIMPLICITY OF RULES}) can be applied upfront to the plan templates. This involves predicting the effectiveness of potential template pairs using the machine learning model, ranking them accordingly, and autonomously selecting the top-$k$ pairs for further processing. Only these top-$k$ template pairs undergo rule enumeration, thereby achieving reduced enumeration time and enabling the enumeration of templates with higher node counts
These templates are instantiated as queries and validated against \textit{standardized rules} to determine their rewrite potential: if a template is deemed rewritable (and thus non-standardized), the generated rules are marked as redundant, and enumeration for that template pair is pruned. 
For valid templates, the workflow proceeds to mapping construction and constraint enumeration, where specific rewriting rules are output.
Then, these valid rules are passed to the RTP stage for redundancy checking.


\textbf{RTP-Based Deduplicator (\S \ref{sec:Rule De-Redundancy}).} 
To mitigate verifying complexity, the RTP algorithm is proposed to systematically prune redundancy while preserving the rule base’s optimization capacity. Theoretically, we ensure that all rules removed are redundant. 
The RTP algorithm initializes an empty rule base for each enumerated template pair. Rules generated in constraint enumeration are first submitted to a checker for verification, followed by redundancy checks for checksum-equivalent rules. If a rule is redundant, RTP halts further enumeration of that rule to reduce subsequent checker calls. Non-redundant rules are included in the rule base, with their constraints evaluated for potential relaxation. Rules that cannot be further constrained are added to the resultant rule set, which is then refined using a multi-way subsumption algorithm.


\textbf{LambdaMART Rule Ranker (\S \ref{Char: SIMPLICITY OF RULES}).} \label{lam}
To identify the most effective rules that maximize system performance, \sysname employs the LambdaMART model~\cite{burges2010ranknet,chapelle2011empirical},  
which is trained to optimize a gradient-boosted decision tree (GBDT) that learns the optimal ranking pattern and further optimize a Normalized Discounted Cumulative Gain (NDCG) that ensures stable ranking of optimal rules. 
This ranker serves a dual purpose: first, it can be optionally applied during the enumeration phase to filter and rank plan template pairs based on predicted effectiveness, selecting only the top-$k$ for rule generation to reduce computational overhead and support enumeration of higher-node templates; second, in the final inference stage, the LambdaMART ranker predicts rule scores from input vectors (formed by tree distance and expression complexity), sorting candidate rules in descending order to obtain an effective rule ranking for query performance optimization.
Note that in contrast to prior studies that rely on brute-force enumeration, which are inherently infeasible for identifying rules in large-node-scale templates (e.g., more than 6 nodes), our ranker achieves two key objectives: (1) it efficiently identifies the most effective rules for small-node-scale templates and (2) systematically explores large-node-scale templates to uncover effective rules that were previously inaccessible. This dual capability not only addresses the scalability limitations of existing approaches but also expands the scope of analyzable templates, thereby enabling comprehensive rule discovery across diverse template sizes.

\end{sloppypar}

%% file: Optimisation-based-on-standardisation.tex
\begin{figure}[!t]
  \centering
  \includegraphics{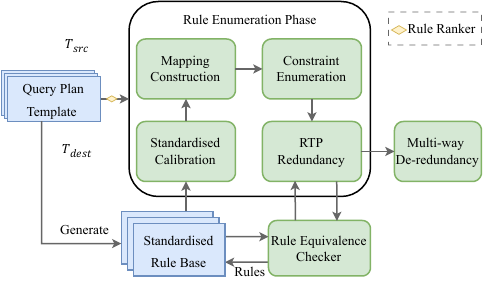}
  \caption{Architecture of standardized rule enumeration.}\label{fig:normalize_structure}
\end{figure}

\begin{sloppypar}

This section outlines the procedure for query rewrite rule enumeration using standardized rules, with the overall architectural workflow illustrated in Figure~\ref{fig:normalize_structure}.
                                                                                                                                                                                                                                                         \sysname first proposes a standardized rule enumeration (\S \ref{sec:Standardisation of rule templates}) that structurally positions projection and selection operators below join operators. This process involves generating a standardized rule base. Notably, this approach also presents theoretically exciting prospects, as enumeration was confined to exponential time verification complexity, while our proposed method enables traceability under certain operations. 
Optionally, to further reduce enumeration overhead and enable enumeration of templates with higher node counts, a template filtering mechanism can be applied using the rule ranker  (\S \ref{Char: SIMPLICITY OF RULES}). This involves ranking potential template pairs based on their predicted effectiveness and selecting only the top-$k$ pairs for rule enumeration, thereby reducing computational complexity and supporting the exploration of complex templates (e.g., with $n \geq 7$ nodes).To further reduce the high overhead of redundancy elimination, we establish Reduced by Template Pair (RTP) optimization (\S \ref{sec:Rule De-Redundancy}), creating a synergistic workflow for efficient rule generation and refinement.

\subsection{Rule Enumerator}
\label{sec:Standardisation of rule templates}
\begin{figure}[t]
  \centering
  \includegraphics[width=0.48\textwidth]{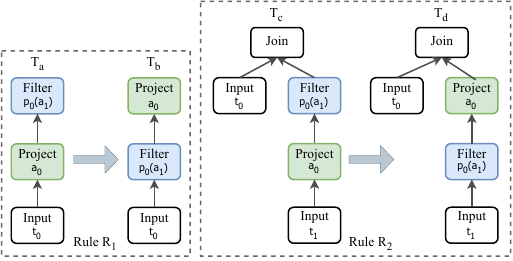}
  \caption{A case for redundancy rules.}\label{fig:normalize_example}
\end{figure}

During rule enumeration, structurally similar templates frequently generate excessive redundant rules. For example, in Figure \ref{fig:normalize_example}, templates  $T_a$ and $T_b$ differ only in the vertical positioning of projection and filter operators, while $T_c$ and $T_d$ are nearly identical except for the swapping of two operators. These trivial structural variations introduce substantial redundancy into the rule set.
To address this, we propose a \textbf{standardized rule} approach to filter out redundant rules at an early stage, minimizing the verification complexity.



\begin{definition}\label{def:standardized_rule}
\begin{sloppypar}
  A standardized rule $R_n(T_{src},T_{dest},M_n,\emptyset)$ is defined by an empty constraint set  $\emptyset$ and a relationship between $T_{src}$ and $T_{dest}$ restricted to two scenarios:
(1). $T_{src}$ and $T_{dest}$ become identical after swapping the positions of two adjacent operators.
  (2). $T_{src}$ and $T_{dest}$ become identical after removing one operator.
  \end{sloppypar}
\end{definition}

Standardized rules uniform query plans into a canonical structure by swapping or removing operators. A key property of these rules is that any non-standardized rule $R(T_{src},T_{dest},M,C)$ is redundant if a standardized rule $R_n(T_{src},T'_{dest},M,\emptyset)$ exists. 

This property enables efficient filtering of redundant rules, which we informally validate through two cases:

\begin{itemize}[leftmargin=*]
    \item For a non-standardized rule $R(T_{src},T'_{dest},M,C)$ where $T_{dest}=T'_{dest}$, the standardized rule \(R_n\) (with an empty constraint set) is more relaxed, making $R$ redundant.
    \item For a non-standardized rule where $T_{dest} \neq T'_{dest}$, any query optimizable by $R$ can also be optimized by \(R_n\) combined with another rule \(R'\), such that $R$ becomes redundant in the set \(\{R_n, R, R'\}\), i.e., _
  \begin{equation}
    \forall q \cdot(\operatorname{Rewrite}(\{R_n,R,R'\}, q)=\operatorname{Rewrite}(\{R_n,R'\}, q))
  \end{equation}
where \(\text{Rewrite}()\) takes a query plan $q$ and a rule set \(\{R\}\) as inputs, and generates an optimized query plan \(q'\) such that the output results of $q$ and \(q'\) are identical.
\end{itemize}

\vspace{1mm}
\noindent \textbf{Case Study.} When employing standardized rules, using Figure \ref{fig:normalize_example} as an illustration, two types of redundant rules can be identify:

(1). Both standardized rule 1 and standardized rule 2 involve similar transformations, with rule 2 being replaced by rule 1. This is due to the fact that Rule 1 contains all necessary conditions to transform the query plan represented by template $T_c$ into $T_d$, making Rule 2 redundant.

(2). Since Rule 1 is a standardized rule, any rule $R_1(T_a, T_e, M, C)$ related to $T_a$ is redundant.

The first type of redundancy can be addressed through de-redundancy after generating the standardized rule base (later in \S \ref{sec:Rule De-Redundancy}). The second type is handled during the rule enumeration.

\vspace{1mm}
\noindent \textbf{Reduce Verifying Overhead for Redundant Templates.} 
Unlike WeTune, our approach reduces verification of redundant rules from exponential to polynomial scale, significantly cutting rule checker verification and lowering enumeration overhead.


Traditional rule enumeration generates all possible template pairs 
$(T_{\text{src}}, T_{\text{dst}})$ and verifies their equivalence, 
with complexity $O(((k^n)^2 \times 2^m)^2)$, where 
$k$ denotes the number of operator types (e.g., JOIN, SELECT), 
$n$ is the number of nodes per template, and $m$ is the number 
of constraint types (e.g., predicates). 
To formalize this, we use first-order logic (FOL) with predicates 
$\text{Template}(T, n)$ ($T$ has $n$ nodes), 
$\text{Equivalent}(T_{\text{src}}, T_{\text{dst}})$ ($T_{\text{src}}$ and $T_{\text{dst}}$ are equivalent), 
and $\text{Enumerate}(P)$ ($P$ is the pair set), we formalize:

\begin{IEEEeqnarray}{c}
\scriptsize
\forall T_{\text{src}}, T_{\text{dst}} \in U \Big(
  \text{Template}(T_{\text{src}}, n) \wedge \text{Template}(T_{\text{dst}}, n) \notag \\
  \quad \rightarrow \exists P \Big(
      P = \{(T_{\text{src}}, T_{\text{dst}})\} \wedge \text{Enumerate}(P)
  \Big)
\end{IEEEeqnarray}

\begin{IEEEeqnarray}{c}
\scriptsize
\forall (T_{\text{src}}, T_{\text{dst}}) \in P \Big(
  \text{Enumerate}(P) \rightarrow \notag \\
  \quad \text{Equivalent}(T_{\text{src}}, T_{\text{dst}}) \vee \neg \text{Equivalent}(T_{\text{src}}, T_{\text{dst}})
\Big)
\end{IEEEeqnarray}

The complexity $O(((k^n)^2 \times 2^m)^2)$ arises from 
$k^n$ templates (each node selects one of $k$ operators), 
$(k^n)^2$ template pairs, and $2^m$ constraint variations 
per template, squared due to pairwise verification.

In contrast, standardized rules and subquery push-up constrain 
template structures, reducing verifiable template pairs from 
$O((k^n)^2)$ to $O(1)$.
We formalize this as:

\begin{IEEEeqnarray}{c}
\scriptsize
\forall T \in U \Big(
  \text{Standardized}(T) \vee \text{PushUpApplicable}(T) \notag \\
  \quad \rightarrow \neg \exists^{\text{many}} (T_{\text{src}}, T_{\text{dst}}) \Big(
      \text{Equivalent}(T_{\text{src}}, T_{\text{dst}}) \wedge \notag \\
      \quad |\{(T_{\text{src}}, T_{\text{dst}})\}| > \text{constant}
  \Big)
\end{IEEEeqnarray}

Empty constraint sets, formalized as:

\begin{IEEEeqnarray}{c}
\scriptsize
\forall T \in U \Big(
  \text{Standardized}(T) \rightarrow \notag \\
  \quad \neg \exists c \in C \, \text{Constraint}(T, c)
\Big)
\end{IEEEeqnarray}

eliminate constraint enumeration, yielding:

\begin{IEEEeqnarray}{c}
\scriptsize
\exists S \subset U \Big(
  \forall T \in S \Big(
      \text{Standardized}(T) \wedge \text{PushUpApplicable}(T)
  \Big) \notag \\
  \quad \rightarrow \text{Complexity}(\text{Optimized}, O(1))
\Big)
\end{IEEEeqnarray}

While this traceability does not apply to all templates:

\begin{IEEEeqnarray}{c}
\scriptsize
\neg \forall T \in U \Big(
  \text{Standardized}(T) \vee \text{PushUpApplicable}(T)
\Big)
\end{IEEEeqnarray}

redundant templates are widespread in benchmarks and real-world scenarios:

\begin{IEEEeqnarray}{c}
\scriptsize
\exists R \subset U \Big( |R| \text{ large} \wedge {} \notag \\ 
\quad \forall T \in R \Big( \text{Redundant}(T) \wedge \text{Standardized}(T) \Big) \Big)
\end{IEEEeqnarray}

Thus, targeted optimizations for these cases yield significant reductions in overall enumeration overhead, validated in benchmarks like TPC-H, achieving up to 90\% time reduction.

\subsubsection{Generation of Standardized Rules}
This part introduces the methodology for generating of standardized rules. Drawing from Definition \ref{def:standardized_rule}, standardized rules are categorized into two types based on their operations:

\begin{itemize}[leftmargin=*]
    \item Operator Removal: These rules simplify query plan templates by eliminating operators that do not impact the final output, thereby streamlining the template structure.
    \item Operator Swapping: These rules adjust the positions of operators and their child nodes to align the query plan template with a canonical structure. 
\end{itemize}

The first type is straightforward: it removes redundant operators to simplify templates. The second type, however, necessitates predefined priorities for different operators to avoid non-convergent transformations. Otherwise, mutual swaps between two operators could prevent the template from stabilizing into a canonical form. To address this, we prioritize three single-input nodes as follows: \text{Project} > \text{Distinct} > \text{Filter}. 







This hierarchy reflects performance impacts analyzed from evaluated queries.
Filter, which does not alter output attribute counts, is prioritized lower and can be moved below Project/Distinct. Project, which modifies attribute lists and influences downstream operations, is pushed upward to maximize applicability. Multi-input operators, less critical for redundancy reduction, are deprioritized. When a parent node has lower priority than its child, a standardized rule swaps their positions for validation.

Multi-input operators are excluded from standardization. Their multiple child nodes complicate post-removal transformation structures, and enforcing priorities would restrict optimization paths—critical for performance, as traditional optimizers already lack sufficient rules for these complex cross-table operations.

Notably, Distinct operators are standardized. Despite their role in optimization, subquery-based Distinct operators are often ineffective (subqueries focus on value existence, not result counts), making their standardization valuable for reducing inefficiencies.

\begin{algorithm}[t]
\small
\caption{Candidate Standardized Rule Generation}\label{algo:normalize_generate}
\KwIn{Template Library $S$}
\KwOut{Candidate Rule Set $NC$}
$NC \gets \emptyset$\;
\ForEach{template $T_{src} \in S$}{
    $T_{dest} \gets \text{clone}(T_{src})$\;
    $M \gets \text{constructNodeMapping}(T_{src}, T_{dest})$\;
    \ForEach{node $n \in T_{dest}$}{
        \eIf{$n$ is a single-input node}{
            Remove $n$ from $T_{dest}$\;
            Update $M$ by removing attributes of $n$\;
            $R_{nc} \gets (T_{src}, T_{dest}, M, \emptyset)$\;
            $NC \gets NC \cup \{R_{nc}\}$\;
            Restore $T_{dest}$ and $M$\;
        }{
            Skip to next node\;
        }
        $n_p \gets \text{parent}(n)$\;
        \If{$n_p \neq \text{null}$ and $\text{priority}(n_p) < \text{priority}(n)$}{
            Swap $n$ and $n_p$ in $T_{dest}$\;
            $R_{nc} \gets (T_{src}, T_{dest}, M, \emptyset)$\;
            $NC \gets NC \cup \{R_{nc}\}$\;
            Restore $T_{dest}$\;
        }
    }
}
\end{algorithm}


Algorithm \ref{algo:normalize_generate} details the generation of candidate standardized rules. For each template \(T_{src}\) in the template library $S$, a copy \(T_{dest}\) is created, and a node mapping $M$ is established between \(T_{src}\) and \(T_{dest}\). The algorithm iterates over all nodes in \(T_{dest}\) to generate rules for operator removal and swapping:
\begin{itemize}[leftmargin=*]
    \item Operator Removal: For each single-input node $n$, $n$ is removed from \(T_{dest}\), and the mapping $M$ is updated by excluding $n$-associated attributes. A new rule \(R_{nc} = (T_{src}, T_{dest}, M, \emptyset)\) is added to the candidate set $NC$, then \(T_{dest}\) and $M$ are restored.
    \item Operator Swapping: For each node $n$ with a parent \(n_p\), if \(n_p\) has lower priority, $n$ and \(n_p\) are swapped in \(T_{dest}\). A new rule is generated and added to $NC$, followed by the restoration of \(T_{dest}\).
\end{itemize}



After generating candidate rules, an equivalence checker validates their correctness, producing a set of valid rules. A deduplication module (details in \S \ref{sec:Rule De-Redundancy}) then removes redundant rules from this set, further enhancing verification efficiency.



\subsubsection{Use of Standardized Rule Base}

  
  
  
  

The standardized rule base is integrated into the rule enumeration framework to optimize query plan templates, as outlined in the following process:
Given a source template \(T_{src}\) and a destination template \(T_{dest}\), \(T_{src}\) is first instantiated into a concrete query plan \(P_{src}\). The optimizer applies rules from the standardized rule base \(S_n\) to transform \(P_{src}\) into an optimized plan set \(P_{dest}\).
\begin{itemize}[leftmargin=*]
    \item If \(P_{dest}\) is non-empty, \(T_{src}\) is deemed redundant as it can be fully optimized using \(S_n\), and no new rules are generated for it.
    \item If \(P_{dest}\) is empty, \(T_{src}\) is irreducible via \(S_n\), and the framework proceeds to enumerate minimal equivalent constraints for the pair \((T_{src}, T_{dest})\), generating non-redundant equivalent rules.
\end{itemize}
This approach, compared to the brute-force state-of-the-art work, efficiently identifies redundant templates and streamlines rule generation, thereby improving the performance of query optimization.


\subsection{Rule Deduplicator}
\label{sec:Rule De-Redundancy}
A key observation is that much redundancy stems from complex mapping constructions and constraint enumerations, which are difficult to filter incrementally during rule generation.
Traditional deduplication, relying on nested loops over rules with \(O(n^2)\) complexity \cite{wang2022wetune}, becomes prohibitively inefficient as template node counts grow: for 4-node templates, de-redundancy exceeds 20 days (Table \ref{tab:normalize_stat}), outpacing enumeration itself due to expanded template complexity and search space. To address this, we propose two optimizations:
(1) partitioning the rule set into smaller subsets, performing de-redundancy on each subset independently, and merging the results for final subsumption checks; (2) introducing the \textbf{Reduce by Template Pair (RTP)} algorithm to enable de-redundancy during the enumeration phase, leveraging template pair properties.

To validate the effectiveness of incremental de-redundancy via subset merging, we briefly prove the following assertion: Given two query rewrite rule sets $S$ and \(S'\) where \(S' \subseteq S\), if a rule \(R \in S'\) is redundant within \(S'\), then R is also redundant within S.

\noindent \textbf{Proof Sketch:} If R is redundant in \(S'\), then by definition:\begin{equation}
\forall q. \left( \operatorname{Rewrite}(S', q) = \operatorname{Rewrite}(S' - \{R\}, q) \right) \label{eq:reduce}
\end{equation}For any query q, optimization using S falls into two cases:If q does not utilize R during optimization, then:
\begin{equation}
\operatorname{Rewrite}(S, q) = \operatorname{Rewrite}(S - \{R\}, q)
\end{equation}If q uses R during optimization, there exists an intermediate query plan \(q_R\) such that:
\begin{equation}
q_R \overset{R}{\rightarrow} q'_R
\end{equation}
By Eq. \ref{eq:reduce}, an alternative optimization path must exist:
\begin{equation}
q_R \overset{R_1}{\rightarrow} \dots \overset{R_n}{\rightarrow} q'_R
\end{equation}
where \(\{R_1, \dots, R_n\} \subseteq S - \{R\}\). Substituting this path into the optimization process yields:
\begin{equation}
\operatorname{Rewrite}(S, q) = \operatorname{Rewrite}(S - {R}, q)
\end{equation}
Thus, R is redundant in S, confirming that incremental subset de-redundancy preserves correctness when merging results.

This property enables efficient redundancy elimination through partitioning. Rules are divided into subsets, processed independently, and merged iteratively to yield a de-redundant set equivalent to processing the entire set at once, accelerating the workflow.

To boost enumeration efficiency, we introduce the Reduce by Template Pair (RTP) algorithm, as outlined in Algorithm \ref{algo:RTP}, which filters redundancy early at the template pair level. For a rule \(R(T_{\text{src}}, T_{\text{dest}}, M, C)\), its template pair is \(P(T_{\text{src}}, T_{\text{dest}})\); after selecting $P$, mapping construction and constraint enumeration proceed with embedded deduplication.



The RTP workflow initializes a dedicated rule base \(S_p\) for each template pair $P$ during enumeration. For each candidate rule $R$ belonging to $P$: If $R$ is non-equivalent, it is marked as NEQ and retained. If $R$ is equivalent, its source template is instantiated into a query plan \(P_{\text{src}}\), which is optimized using \(S_p\) and \(S_p \cup \{R\}\) respectively. If the optimization results are identical, $R$ is redundant (marked as Repeat) and discarded; otherwise, $R$ is non-redundant (marked as EQ) and added to \(S_p\). 

\begin{algorithm}[t]
  \SetAlgoLined
  \normalem
  \small
  \caption{Reduce by Template Pair (RTP)}\label{algo:RTP}
  \KwIn{Candidate Rule $R$, De-redundant Rule Base $S_p$}
  \KwOut{Status of rule in current rule base (EQ, NEQ, or Repeat)}
  
  \If{rule $R$ is not equivalent}{
    \Return{NEQ}\;
  }
  
  $P_{src} \gets \text{instantiate Source Template for rule } R$\;
  
  The optimiser uses the $S_p$ rule base to optimise $P_{src}$\;
  
  $Ps_{dest} \gets \text{optimise } P_{src} \text{ using rule base } S_p$\;
  
  $S'_p \gets S_p \cup \{R\}$\;
  
  $Ps'_{dest} \gets \text{optimise } P_{src} \text{ using rule base } S'_p$\;
  
  \If{$Ps_{dest} = Ps'_{dest}$}{
    \Return{Repeat}\;
  }
  
  \Return{EQ}\;
\end{algorithm}

\end{sloppypar}

%% file: Simplicity-of-rules.tex
\begin{figure*}[!htbp]
  \centering
  \includegraphics{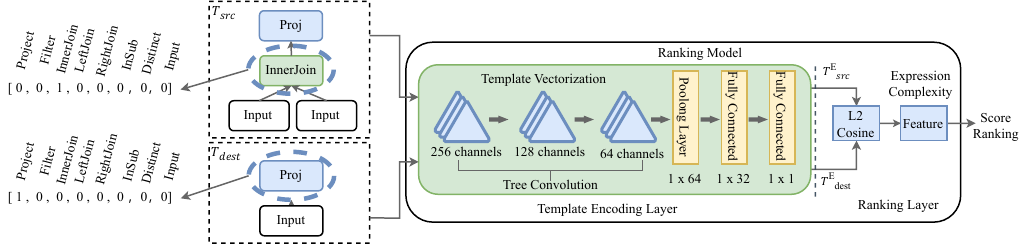}
  \caption{Architecture of rule effectiveness ranking.}\label{fig:Rule Effectiveness Ranking}
\end{figure*}

This section elaborates on the application of the learning ranking model in the effectiveness ranking of rules and templates. Its architectural workflow is shown in Figure ~\ref{fig:Rule Effectiveness Ranking}.
We adopt a LambdaMART model to leverage its capability in handling large-scale datasets and optimizing ranking performance. It covers model design (see \S \ref{sec: Model Design} for details), training (see \S \ref{sec: Model Training} for details) procedures, and applications to rule enumeration (see \S \ref{sec: Model Using} for details).


\subsection{Model Design}\label{sec: Model Design}

\begin{sloppypar}

The design of the LambdaMART model for rule effectiveness ranking consists of Template Encoding and Ranking layers, as depicted in Figure~\ref{fig:Rule Effectiveness Ranking}.
This part explains the methods (i.e., encoding process, ranking model, and loss function) employed in these two layers. 

\subsubsection{Encoding Process}\label{sec: Encoding Process}

\begin{table}[!t]
\centering
\caption{The formal definitions of operators based on probability distribution functions.}
\label{tab:operator_defs}
\footnotesize 
\setlength{\tabcolsep}{3pt} 
\renewcommand{\arraystretch}{1.5}

\resizebox{\columnwidth}{!}{%
  \begin{tabular}{cc}
      \hline
      \textbf{Operator} & \textbf{Expression} \\
      \hline
      $\mathrm{Input}_r$ & $f(t) := r(t)$ \\
      \hline
      $\mathrm{Proj}_a$ & $f(t) := \sum_{t_l} (f_l(t_l) \times [\![t = a(t_l)]\!])$ \\
      \hline
      $\mathrm{Sel}_{p, a}$ & $f(t) := f_l(t) \times [\![p(a(t))]\!]$ \\
      \hline
      $\mathrm{InSubSel}_a$ & $f(t) := f_l(t) \times ||f_r(a(t))|| \times \mathrm{not}([\![\mathrm{IsNull}(a(t))]\!])$ \\
      \hline
      $\mathrm{IJoin}_{a_l, a_r}$ & 
      $\begin{gathered} 
      f(t) := \sum_{t_l, t_r} ([\![t = t_l \cdot t_r]\!] \times f_l(t_l) \times f_r(t_r) \times [\![a_l(t_l) = a_r(t_r)]\!] \\ 
      \times \mathrm{not}([\![\mathrm{IsNull}(a_l(t_l))]\!])) 
      \end{gathered}$ \\
      \hline
      $\mathrm{LJoin}_{a_l, a_r}$ & 
      $\begin{gathered} 
      f(t) := (\mathrm{IJoin\;Expr}) + \sum_{t_l, t_r} ([\![t = t_l \cdot t_r]\!] \times f_l(t_l) \times [\![\mathrm{IsNull}(t_r)]\!] \\
      \times \mathrm{not}(\sum_{t'_r} (f_r(t'_r) \times [\![a_l(t_l) = a_r(t'_r)]\!] \times \mathrm{not}([\![\mathrm{IsNull}(a_l(t_l))]\!])))) 
      \end{gathered}$ \\
      \hline
      $\mathrm{RJoin}_{a_l, a_r}$ & \text{(symmetric to LJoin)} \\
      \hline
      $\mathrm{Dedup}$ & $f(t) := ||f_l(t)||$ \\
      \hline
  \end{tabular}%
} 
\end{table}

The planar graph embedding layer in the ranking model maps each rule \( R = (T_{\text{src}}, T_{\text{dest}}, M, C) \), comprising the source template \( T_{\text{src}} \) and destination template \( T_{\text{dest}} \), from the original feature space to a nine-dimensional (9-dim) embedding space, in order to learn differences between plans. This mapping captures the differences between tree-structured query plans. The template vectorization process extracts features from each query plan tree to generate its corresponding embedding. For tree-structured query plans, designing an embedding model that captures critical structural information is crucial for ensuring the model's efficacy and efficiency.


\sysname builds upon a tree convolutional model, drawing inspiration from \cite{marcus2021bao, marcus2019neo, mou2016convolutional, zhu2023lero}, but incorporates significant enhancements, including omitting estimated costs and focusing solely on node properties. As shown in Figure~\ref{fig:Rule Effectiveness Ranking}, it converts tree-structured query plans into vectorized representations where nodes are one-hot encoded by operation type, and convolution captures structural relationships.
Unlike \cite{marcus2021bao, marcus2019neo, zhu2023lero}, this representation excludes real-world conditions related to tables, cardinalities, or other external factors, concentrating exclusively on the intrinsic properties of the tree nodes themselves. The tree convolution applies multiple convolutional kernels over each node and its two child nodes, facilitating the transformation of one plan tree into another. Subsequently, the node vectors are aggregated via dynamic pooling and processed by the neural network to produce the 9-dim embedding. Further operations are applied to the vector representations of \( T_{\text{src}} \) and \( T_{\text{dest}} \) obtained from the template encoding layer to generate a suitable feature vector.

The encoding phase converts raw query rewrite rules into a feature-rich representation amenable to machine learning. This stage produces a feature vector encompassing nine attributes:

\textbf{L2 Distance}: Measures the \textcircled{1} structural transformation magnitude between $T_{\text{src}}$ and $T_{\text{dest}}$, computed as:
  \begin{equation}
  \text{L2 Distance} = \sqrt{\sum_{i=1}^{n} (e_{\text{src},i} - e_{\text{dest},i})^2},
  \end{equation}
  where $e_{\text{src},i}$ and $e_{\text{dest},i}$ are components of tree embedding vectors.
  
\textbf{Cosine Similarity}: Assesses the \textcircled{2} angular similarity between template embeddings:
  \begin{equation}
  \text{Cosine Similarity} = \frac{\sum_{i=1}^{n} e_{\text{src},i} \cdot e_{\text{dest},i}}{\sqrt{\sum_{i=1}^{n} e_{\text{src},i}^2} \cdot \sqrt{\sum_{i=1}^{n} e_{\text{dest},i}^2}}.
  \end{equation}

  \textbf{Expression Complexity Features}: These include metrics derived from operator expressions in Table~\ref{tab:operator_defs}, averaged over primary operators, capturing syntactic and semantic intricacy: \textcircled{3} $\text{expr\_length}$ as the string length of the expression, indicating syntactic volume; \textcircled{4} $\text{sum\_count}$ as the count of $\sum$ operators, quantifying aggregation depth; \textcircled{5} $\text{times\_count}$ as the count of $\times$ operators, quantifying multiplicative dependencies; \textcircled{6} $\text{neg\_count}$ as the count of $\neg$ operators, quantifying logical negation; \textcircled{7} $\text{bracket\_count}$ as the number of [ ] bracket pairs, quantifying conditional branching; \textcircled{8} $\text{var\_count}$ as the number of variables (e.g., t, r, f, a), refined as $\max(\text{raw count}, \text{<null> count})$ to handle abstract templates; and \textcircled{9} $\text{complexity}$ as the total operator count (sum + times + neg + bracket), serving as a proxy for verification overhead.
  


Tree embeddings are derived using a neural network that processes the abstract syntax tree (AST) of each template, capturing hierarchical relationships.

\subsubsection{Ranking Model} \label{sec: Ranking Model}

The ranking model employs Gradient Boosted Decision Trees (GBDT) \cite{friedman2001greedy}, a machine learning algorithm that iteratively constructs a set of decision trees to enhance predictive accuracy. It belongs to the boosting algorithm family, where each tree is trained to correct the residuals (prediction errors) of previous trees. The process uses gradient descent to minimize a specified loss function, with each new tree added along the negative gradient direction. The final prediction is a weighted sum of all trees, adjusted by a learning rate to prevent overfitting. 

In the context of \sysname, GBDT is implemented using the LightGBM library, which optimizes the LambdaMART algorithm for scoring and ranking tasks by adjusting tree parameters (e.g., the number of leaves and minimum data points per leaf). The process begins with the 9-dim embedding vectors of \( T_{\text{src}} \) and \( T_{\text{dest}} \), derived from the Tree Convolutional Neural Network (TCNN) as described in \S \ref{sec: Encoding Process}. This feature set is incorporated into a training dataset via a function that constructs a LightGBM Dataset, paired with relevance labels from the corresponding data.

The \sysname model employs a Gradient Boosted Decision Tree (GBDT) framework for training, with parameters including a learning rate of 0.1, 62 leaves per tree, a minimum of 20 data points per leaf, and 100 boosting iterations, generating a rule score \( s_k \) as the weighted sum of predictions from 100 decision trees. The learning rate of 0.1 balances convergence and overfitting prevention, while 62 leaves per tree, determined through cross-validation, effectively capture the detailed features of the rules. A minimum of 20 data points per leaf mitigates overfitting, and 100 iterations ensure convergence while optimizing the Normalized Discounted Cumulative Gain (NDCG@k), achieving a score exceeding 0.9. These scores, denoted as \( s_k \), are computed as the weighted sum of predictions from the 100 decision trees: $s_k = \sum_{t=1}^{T} f_t(F_k)$
where \( F_k \) represents the feature vector of rule \( R_k \), and \( f_t \) is the prediction of the \( t \)-th tree. The score \( s_k \) reflects the rule's relative effectiveness, with higher values indicating greater potential to optimize query execution. The trained model is saved to and applied in the function for inference.

\subsubsection{Loss Function}

The ranking model aims to maximize the likelihood of outputting the correct order within the rule set to select the best top plans; thus, the loss function for the ranking model is designed to achieve this goal. The loss function utilizes the  $lambdarank $ objective implemented in LightGBM to optimize the ranking model. This loss function is intended to maximize the Normalized Discounted Cumulative Gain at \( n \) (NDCG@\( n \)), a ranking metric used to evaluate the quality of the top \( n \) ranked rules.

NDCG@\( n \) focuses on the top \( n \) positions, with the calculation formula as follows: $\text{NDCG@}n = \frac{\text{DCG@}n}{\text{IDCG@}n}$
where \(\text{DCG@}n = \sum_{i=1}^{n} \frac{2^{\text{rel}_i} - 1}{\log_2 (i+1)}\) is the Discounted Cumulative Gain, \(\text{rel}_i\) is the relevance score (1 indicating a rule reduces execution time, 0 indicating an ineffective rule), and \(\log_2 (i + 1)\) is a position-based discount factor. \(\text{IDCG@}n\) is the maximum possible DCG for the top \( n \) ranks, computed using the optimally ordered relevance scores.

LambdaRank: The loss is determined by the Lambda weight (\(\lambda_{ij}\)), which reflects the change in NDCG when the ranks of two items are swapped.

Lambda weight (\(\lambda_{ij}\)): A weighting factor that measures the impact of swapping the ranks of two rules \( R_i \) and \( R_j \) on the NDCG@\( n \) metric, defined as:
$\lambda_{ij} = \frac{\partial \text{NDCG@}n}{\partial s_i} - \frac{\partial \text{NDCG@}n}{\partial s_j}$
where \(\frac{\partial \text{NDCG@}n}{\partial s_k}\) is the gradient of NDCG@\( n \) with respect to the score \( s_k \) of rule \( R_k \), approximated based on the relevance difference and position discount.
Rule pairs are adjusted in their relative order to ensure that rules with higher relevance scores are ranked above those with lower scores.

Relevance score (\(\text{rel}_i\)): A binary label indicating a rule’s effectiveness, derived from execution time reduction, where 1 denotes a beneficial rule and 0 denotes a non-beneficial one.

The loss function in LambdaMART is based on a pairwise ranking approach, aiming to minimize the difference between the predicted score ordering and the ground truth relevance ordering. Specifically, for any pair of rules \((R_i, R_j)\) with relevance scores \(\text{rel}_i\) and \(\text{rel}_j\), the model adjusts the scores \( s_i \) and \( s_j \) such that if \(\text{rel}_i > \text{rel}_j\), then \( s_i > s_j \). The loss is computed using a sigmoid-based pairwise error term, weighted by \(\lambda_{ij}\), and optimized via gradient descent within the GBDT framework.

\subsection{Model Training} \label{sec: Model Training}

Unlike conventional approaches that depend on pre-trained cost models or synthetic workloads, our training uses real-world SQL transaction data, focusing solely on intrinsic rewrite rule differences (excluding external factors like table data, cardinality, or estimated costs). This part outlines the pipeline for real transaction workloads: templatized SQL statements (matching rule source templates) are rewritten using applicable rules, with effectiveness validated via runtime comparisons in a production database. Rules reducing latency are labeled 1, ineffective ones 0, ensuring the model learns from practical outcomes rather than synthetic estimates.



The feature set is extracted solely from rule representations, as detailed in \S \ref{sec: Encoding Process}. These features are integrated into a LightGBM Dataset with a single group \([len(X_{train})]\), reflecting a unified query context, which supports the \(lambdarank\) objective for pairwise rule optimization based on relevance labels.

Training employs the LambdaMART algorithm within LightGBM, configured to rank the top \(k\) beneficial rules effectively. The process iterates over 100 boosting rounds, constructing an ensemble of decision trees to adjust predicted scores \(s_k = \sum_{t=1}^{100} f_t(F_k)\), where \(F_k\) is the feature vector of rule \(R_k\) and \(f_t\) is the \(t\)-th tree's prediction. The \(lambdarank\) loss function maximizes Normalized Discounted Cumulative Gain (NDCG@\(k\)), as defined in the mathematical framework of \S \ref{sec: Ranking Model}, with the Lambda weight \(\lambda_{ij}\), detailed in the same section, guiding gradient descent to enforce correct ordering. Upon completion, the model ranks rules adaptively for specific workloads, enhancing efficiency without reliance on external cost models.

\subsection{Applications To Rule Enumeration} \label{sec: Model Using}

To enhance the efficiency of rule discovery, particularly for templates with higher node counts, the LambdaMART ranker can be optionally applied during the template enumeration phase as a pre-filtering mechanism. This approach leverages the ranker's predictive capabilities to prioritize template pairs based on their estimated effectiveness in yielding performance-improving rules.
Specifically, prior to full rule enumeration, the ranker evaluates potential source-destination template pairs by generating feature vectors derived from template structures, such as operator compositions. These vectors are fed into the trained LambdaMART model, which outputs predicted scores reflecting the likelihood of deriving effective rewrite rules from each pair. The template pairs are then sorted in descending order of these scores, and only the top-$k$ pairs are selected for subsequent processing, including instantiation, validation, and constraint enumeration.

This selective filtering introduces a configurable trade-off between computational efficiency and comprehensiveness. Larger values of k increase the enumeration time due to the increased number of candidate pairs to be processed, but they also increase the fraction of valid rules captured relative to a complete (brute force) enumeration of all possible pairs. Conversely, a smaller $k$ minimizes overhead, enabling scalable exploration of large-node templates (e.g., $n \geq 7$) that would otherwise be infeasible, at the cost of potentially missing some lower-ranked but still valuable rules. Users can autonomously tune $k$ based on resource constraints and desired coverage, thereby balancing enumeration latency with the effectiveness ratio of the resultant rule set. Empirical tuning of $k$ is recommended to optimize this balance for specific database workloads.

\end{sloppypar}

%% file: Evaluation.tex
\begin{sloppypar}

The evaluation aims to answer the following questions (Q):



\begin{itemize}[leftmargin=*]
\item \textbf{Q1}: Compared to SOTA WeTune, what is the efficiency of \sysname in enumerating effective rules? Specifically, does it demonstrate fast performance on small-scale tasks, and is it feasible to apply it to large-scale scenarios? (\S \ref{sec:evaluation_efficiency})

\item \textbf{Q2}: Compared to SOTA WeTune, what is the effectiveness of \sysname in enumerating rules? This includes aspects such as whether it can enumerate a greater number of rules and effective rules that were not reported previously. (\S \ref{sec:evaluation_effectiveness})

\item \textbf{Q3}: What are the individual and combined contributions of \sysname’s proposed components to its overall efficiency and effectiveness? This details the ablation study of our proposed rule enumerator, deduplicator, and ranker. (\S \ref{sec:evaluation_ablation_study})
\end{itemize}

\subsection{Experimental Setup}

\textbf{Testbed.} 
The experiments were conducted on a server equipped with an Intel(R) Xeon(R) Silver 4216 processor (64x 2.10 GHz), featuring 16 physical cores and 32 threads, with 376 GB of memory. The thread count was set to 16 to minimize context-switching overhead, thereby optimizing performance for both rule generation and model training. SQL
Server 2019 \cite{sqlserver2019} is deployed for model training of \sysname's ranker.


\vspace{1mm}
\noindent \textbf{Baseline.} 
To address RQ1 and RQ2, \sysname is primarily compared against SOTA WeTune’s \cite{wang2022wetune}. The comparison focuses on efficiency, with key metrics including enumeration time, verification calls, the maximum number of nodes in templates, and the improvement in identifying useful rules. For RQ3, we conducted an ablation study on our proposed rule enumerator (\S \ref{sec:Standardisation of rule templates}), deduplicator (\S \ref{sec:Rule De-Redundancy}), and ranker (\S \ref{Char: SIMPLICITY OF RULES}).

\vspace{1mm}

\noindent \textbf{Generating Rules.}
\sysname employs a template-optimized enumeration algorithm to generate query plan templates with 5 and 6 nodes, resulting in 26,508 and 652,808 new non-redundant and high-potential rules, respectively. In total, 679,316 new rules are added to the rule base. In general, WeTune is infeasible to enumerate rules for templates with more than 4 nodes. 

\vspace{1mm}

\noindent \textbf{Evaluating Rules.} 
The optimization performance of the LambdaMART model is assessed by rewriting real-world SQL queries using rule libraries both before and after ranking, followed by quantifying the ranking positions of rules employed during actual SQL execution. The model is trained on rules validated within a SQL Server environment, where rules that reduce execution time are assigned a relevance score of 1. We compare the model’s performance against a baseline of random rule ordering, employing the NDCG@k metric (optimized during training via the LambdaRank objective function) to quantify the accuracy of ranking effective rules in top positions. This evaluation approach ensures that the model prioritizes rules with discussed optimization potential, which aligns with the goal of enhancing query performance.

\vspace{1mm}

\noindent \textbf{Datasets/Queries Statistics.} We evaluate the \sysname model using a custom database workload, building upon the workload methodology consistent with WeTune \cite{wang2022wetune}, which includes real-world SQL queries. These queries are collected from 15 open-source web applications hosted on GitHub, selected based on their popularity (star counts ranging from 1,500 to 20,000) and encompassing diverse genres such as e-commerce, content management, discussion forums, and social networking. These applications, contributed by 1 to 1,800 developers, yield 8,518 unique SQL queries extracted through unit tests executed on their source code as of July 2025. To complement open-source queries, we further incorporate 3,000 commercial queries from Yashan DB~\cite{yashandb}, a production-ready database system, featuring intricate relational queries tailored to financial and internal management systems. The combined workload of 11,518 queries is validated by parsing and analyzing their structural features (e.g., join counts, subquery depths), ensuring compatibility with the rule optimization framework.


\begin{figure}[!t]
  \centering
    \begin{subfigure}[t]{0.49\columnwidth}
    \centering
    \includegraphics[width=\linewidth]{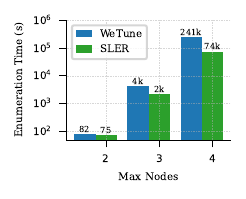}
    \caption{Enumeration time}
    \label{fig:enumeration_time}
  \end{subfigure}
  \begin{subfigure}[t]{0.49\columnwidth}
    \centering
    \includegraphics[width=\linewidth]{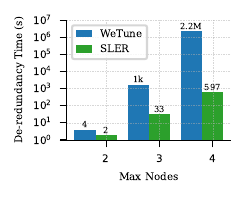}
    \caption{\mbox{De-redundancy time}}
    \label{fig:checker_calls}
  \end{subfigure}
  \caption{Efficiency evaluation of \sysname (log scale).}
  \label{fig:evaluation_SOTA_comparison}
\end{figure}

\subsection{Efficiency}\label{sec:evaluation_efficiency}
This part compares \sysname to SOTA WeTune in terms of enumeration and de-redundancy time. \sysname beats WeTune in both small-scale and large-scale rule enumeration, with faster execution for small-node rules and feasible scalability for large-node scenarios.

\noindent \textbf{Small-Scale Efficiency.} For rules with 2–4 nodes, \sysname reduces enumeration time significantly. As shown in Figure \ref{fig:evaluation_SOTA_comparison}, 
\sysname completes 2-, 3-, and 4-node enumeration in 72 seconds (87.8\% of WeTune’s time), 2,362 seconds (55.14\% of WeTune’s time), and 82,763 seconds (34.27\% of WeTune’s time), respectively. 
Also, \sysname completes 2-, 3-, and 4-node de-redundancy in 2 seconds (50\% of WeTune’s time), 33 seconds (2\% of WeTune’s time), and 597 seconds (<1\% of WeTune’s time), respectively. 


\begin{table}[!t]
  \caption{Efficiency evaluation of \sysname compared to WeTune in enumeration and de-redundancy.
  A ``timeout'' is an estimated runtime exceeding 10 years, signifying scenarios where rule enumeration or verification becomes computationally infeasible.
  }\label{tab:normalize_stat}
  \belowrulesep=0pt
  \aboverulesep=0pt
  \centering
  \footnotesize 
  \renewcommand{\arraystretch}{1.2}
  
  \resizebox{\columnwidth}{!}{%
    \begin{tabular}{ccccccc} 
      \toprule
      \multicolumn{2}{c}{Max nodes in the template} & \multicolumn{1}{c}{2} & \multicolumn{1}{c}{3} & \multicolumn{1}{c}{4} & \multicolumn{1}{c}{5} & \multicolumn{1}{c}{6}\\
      \midrule
      \multicolumn{1}{c|}{\multirow{3}{*}{
        \begin{tabular}{@{}c@{}}Enumeration \\ Duration\end{tabular}
      }} & WeTune & 82s & 1h 11m & 67h 4m & timeout & timeout \\
      \multicolumn{1}{c|}{} & \begin{tabular}{@{}c@{}}\sysname \\ (w.o. RTP)\end{tabular} & 72s & 39m 22s & 22h 59m & timeout & timeout \\
      \multicolumn{1}{c|}{} & \sysname & 75s & 36m 46s & 20h 49m & 29d 23h & 395d \\ \hline
      \multicolumn{1}{c|}{\multirow{3}{*}{
        \begin{tabular}{@{}c@{}}De-redundancy \\ Duration\end{tabular}
      }} & WeTune & 4s & 28m 53s & 20d 19h & timeout & timeout \\
      \multicolumn{1}{c|}{} & \begin{tabular}{@{}c@{}}\sysname \\ (w.o. RTP)\end{tabular} & 3s & 2m 27s & 33m 33s & timeout & timeout \\
      \multicolumn{1}{c|}{} & \sysname & 2s & 33s & 9m 57s & 10h 24m & 62h 24m \\
      \bottomrule
    \end{tabular}%
  } 
\end{table}

\noindent \textbf{Large-Scale Feasibility.} For rules involving 5 or more nodes, as shown in Table \ref{tab:normalize_stat}, WeTune fails to enumerate rules (timeout), whereas \sysname is feasible to complete 5-node enumeration in 30 days and 6-node in 395 days. While specific timelines for 7-node and larger templates remain unquantified, the framework’s ranker design enables scalable enumeration for such cases through an effectiveness-driven rule sorting strategy. This strategy leverages specific SQL template features and preliminary LambdaMART training to pre-sort rules, prioritizing the top $n$ rules (where $n$ is dynamically determined based on NDCG@k confidence intervals) for subsequent validation and redundancy elimination. By reducing validation overhead through this targeted prioritization, \sysname possibly facilitates scalable (but incomplete) enumeration for rules with 7 or more nodes (more details in \S \ref{sec:discussion}).

\noindent \textbf{Large-Scale Feasibility.} 
WeTune fails to enumerate rules involving $n \geq 5$ (timeout), and even \sysname takes 395 days for $n=6$ (Table \ref{tab:normalize_stat}). To enable scalable enumeration for $n \geq 7$ templates, the framework leverages its effectiveness-driven  ranker design . This strategy uses SQL template features and LambdaMART training to  pre-sort  rules, prioritizing the  top $n$ rules  for validation (where $n$ is dynamically determined). This targeted prioritization significantly reduces validation overhead, making  scalable (but incomplete) enumeration  for $n \geq 7$ nodes feasible. 

Figure \ref{fig:template_filtering_analysis_a} highlights an optimal efficiency trade-off: a 40\% template filtering rate yields over 80\% effective rule coverage in just \textbf{143 minutes}. This configuration offers high efficiency under a controlled trade-off, making it a practical choice for scalability. Figure \ref{fig:template_filtering_analysis_b} further demonstrates \sysname's scalability to complex 10-node rules. When scaling up to 5k templates, the system discovers over 5.1k effective rules in 523 minutes, confirming its strong scalability even for significantly larger rule templates.


\sysname's small-scale efficiency and large-scale feasibility derive from standardized templates and RTP, which reduce redundancy and verification overhead, critical as node counts and redundant rules grow.

\subsection{Effectiveness}\label{sec:evaluation_effectiveness}
This part compares \sysname to SOTA WeTune in terms of the number and effectiveness of rules.
\sysname demonstrates superior effectiveness by enumerating more rules, including large-node rules, and effective rules that were not reported previously. 

\noindent \textbf{Rule Quantity and Scalability.} WeTune is limited to 4-node rules, while \sysname generates 26,508 and 652,808 new rules for 5 and 6 nodes, expanding the rule base to 679,316 rules, which is 98.25x larger than WeTune’s 4-node rule set. These rules cover previously unexplored patterns, enabling optimization of complex queries. For rules involving 7 or more nodes, where full enumeration is computationally infeasible, \sysname employs an effectiveness-driven template filtering strategy using the LambdaMART ranker (\S \ref{Char: SIMPLICITY OF RULES}). By controlling the number of top-$k$ template pairs selected for enumeration, \sysname facilitates the generation of higher-node rules, balancing enumeration time and rule effectiveness to explore previously inaccessible query patterns.

\begin{figure}[!t]
  \centering
  \begin{subfigure}[t]{0.49\columnwidth}
    \centering
    \includegraphics[width=\linewidth]{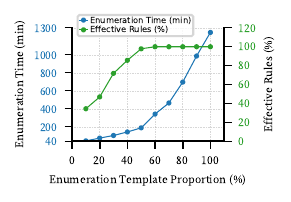}
    \caption{Template filtering with 4-node rules}
    \label{fig:template_filtering_analysis_a}
  \end{subfigure}
  \hfill 
  \begin{subfigure}[t]{0.49\columnwidth}
    \centering
    \includegraphics[width=\linewidth]{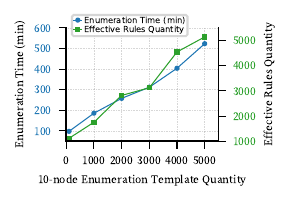}
    \caption{Template filtering of 10-node rules}
    \label{fig:template_filtering_analysis_b}
  \end{subfigure}
  \caption{Analysis of Rule Enumeration Efficiency and Effectiveness Across Different Template Proportions and Node Counts.}
  \label{fig:template_filtering_analysis}
\end{figure}

As shown in the Figure \ref{fig:template_filtering_analysis}, it is noteworthy that with a template filtering rate of 40\%, \sysname achieves over 80\% effective rule coverage with an enumeration time of merely 143 minutes. This configuration offers higher efficiency under a controlled trade-off in effective rule proportion, making it a practical choice for scalability.

\begin{table}[t!]
  \caption{Representative examples of standardized rules. $\mathrm{Proj}$, $\mathrm{IJoin}$, $\mathrm{LJoin}$, $\mathrm{{Proj}^*}$, and $\mathrm{InSub}$ stand for a $\mathrm{predicate}$, $\mathrm{InnerJoin}$, $\mathrm{LeftJoin}$, $\mathrm{Distinct}$, and $\mathrm{SubQuery}$, respectively.}\label{tab:normalize_example}
  \belowrulesep=0pt
  \aboverulesep=0pt
  \centering
  \footnotesize
  \setlength{\tabcolsep}{4pt}
  
  \resizebox{\columnwidth}{!}{%
    \begin{tabular}{c | c | c}
      \toprule
      No. & \thead{Source Plan \\ Template} & \thead{Destination Plan \\ Template} \\ 
      \midrule
      1 & $\mathrm{Filter_{p_0,a_1}(Proj_{a_0}(t_0))}$ & $\mathrm{Proj_{a_0}(Filter_{p_0,a_1}(t_0))}$ \\ 
      2 & $\mathrm{Filter_{p_0,a_1}({Proj}^*_{a_0}(t_0))}$ & $\mathrm{{Proj}^*_{a_0}(Filter_{p_0,a_1}(t_0))}$ \\ 
      3 & $\mathrm{Proj_{a_1}(Proj_{a_0}(t_0))}$ & $\mathrm{Proj_{a_1}(t_0)}$ \\ 
      4 & $\mathrm{{Proj}^*_{a_1}({Proj}^*_{a_0}(t_0))}$ & $\mathrm{{Proj}^*_{a_1}(t_0)}$ \\ 
      5 & $\mathrm{Proj_{a_3}(IJoin_{a_1,a_2}(Proj_{a_0}(t_0),t_1))}$ & $\mathrm{Proj_{a_3}(IJoin_{a_1,a_2}(t_0,t_1))}$ \\ 
      6 & $\mathrm{Proj_{a_3}(LJoin_{a_1,a_2}(Proj_{a_0}(t_0),t_1))}$ & $\mathrm{Proj_{a_3}(LJoin_{a_1,a_2}(t_0,t_1))}$ \\ 
      7 & $\mathrm{InSub_{a_1,a_2}(Proj_{a_0}(t_0),t_1)}$ & $\mathrm{Proj_{a_0}(InSub_{a_1,a_2}(t_0,t_1))}$ \\ 
      8 & $\mathrm{InSub_{a_1,a_2}({Proj}^*_{a_0}(t_0),t_1)}$ & $\mathrm{{Proj}^*_{a_0}(InSub_{a_1,a_2}(t_0,t_1))}$ \\ 
      \bottomrule
    \end{tabular}%
  } 
\end{table}

Table \ref{tab:normalize_example} presents a representative subset of 8 standardized rules, encompassing nearly all rule types. Rules 1 and 2 belong to the operator exchange category, while rules 3 and 4 are operator deletion rules. Rules 5–6 focus on join operations, and rules 7–8 target subquery transformations.
This set is representative due to its coverage of diverse query rewriting scenarios. As detailed in \S \ref{sec:Standardisation of rule templates}, these standardized rules eliminate redundancies by leveraging structural similarity while preserving semantically distinct variants.


\noindent \textbf{Rule Quality.} \sysname’s rules enable direct, one-step optimization of queries, where WeTune requires multiple steps. For example in Table \ref{tab: example of query}, WeTune fails to rewrite query \(q_1\) (no applicable rules). Query \(q_4\)  is optimized to \(q_6\) using a single 5-node rule from \sysname (Rule 2 from Table \ref{tab: example of rules}), whereas WeTune needs four steps of rule applications. 
For query \(q_7\), WeTune can only produce a non-minimal form. This stems from its primary design for query patterns with four or fewer nodes. Also, Table \ref{tab: example of rules} shows more new rules, none of which are discovered by the WeTune.


\begin{table}[!t]
  \caption{Evaluation of \sysname's Enumerator (OPT1 in \S\ref{sec:Standardisation of rule templates}) and Deduplicator (OPT2 in \S\ref{sec:Rule De-Redundancy}) in terms of the number of equivalence verification calls and enumeration time.}
  \label{tab:normalize_callbrator}
  \centering
  \footnotesize
  \setlength{\tabcolsep}{2.5pt} 
  \renewcommand{\arraystretch}{1.1}
  \renewcommand{\theadfont}{\footnotesize} 

  \resizebox{\columnwidth}{!}{%
    \begin{tabular}{c*{6}{c}}
      \toprule
      \multirow{2}{*}{\thead{Max \\ Nodes}}
      & \multicolumn{3}{c}{\thead{\#Verification \\ Calls}} 
      & \multicolumn{3}{c}{\thead{Enumeration \\ Time (s)}} \\
      \cmidrule(lr){2-4} \cmidrule(lr){5-7}
      & No OPT & OPT1 & OPT1+OPT2 & No OPT & OPT1 & OPT1+OPT2 \\
      \midrule
      2 & 24,015 & 48.23\% & 45.98\% & 82 & 87.80\% & 91.46\% \\
      3 & 5,839,699 & 49.76\% & 45.74\% & 4,284 & 55.14\% & 51.48\% \\
      4 & 314,860,926 & 31.99\% & 28.97\% & 241,457 & 34.27\% & 31.05\% \\
      \bottomrule
    \end{tabular}%
  } 
\end{table}

\begin{table}[!t]
  \caption{Standardized rule generation overhead.}\label{tab:normalize_rules}
  \centering
  \footnotesize
  \setlength{\tabcolsep}{4pt}
  \renewcommand{\arraystretch}{1.2}
  \resizebox{0.47\textwidth}{!}{
  \begin{tabular}{cccccc}
    \toprule
    Max Nodes in the template & 2     & 3     & 4    & 5  & 6\\
    \midrule
    Enumeration overhead     & 1.47s & 4.69s & 33.20s &5m 42s  &38m 44s  \\
    De-redundancy overhead    & 2.10s & 2.93s & 4.45s  &2m 44s  &16m 28s\\
    Total overhead      & 3.57s & 7.62s & 37.65s &8m 26s  &55m 12s   \\
    Overhead ratio     & 4.64\% & 0.34\% & 0.0498\% & 0.0193\%  & 0.0104\%  \\
    Number of standardized rules  & 5     & 25    & 77     &204  &1030  \\
    \bottomrule
  \end{tabular}
  }
\end{table}

\begin{table}[t]
\centering
\caption{Evaluation of \sysname's ranker in terms of trigger rate and rule usage. 
}
\label{tab:rule_effectiveness}
\footnotesize	
\resizebox{0.48\textwidth}{!}{
    \begin{tabular}{ccccc}
    \toprule
    Rule Top  & Rule End  & Trigger Rate (\%) & Top Rule Usage(\%) & End Rule Usage(\%) \\
    \midrule
     Top 50 & End 150 & 100 & 52 & 0 \\
     Top 500 & End 200 & 100 & 34.67 & 0 \\
     Top 1000 & End 0 & 100 & 11.4 & 0 \\
     Top 0 & End 1000 & 0 & 0 & 0 \\
    \midrule 
    \multicolumn{2}{c}{Random 1000} & 0& 0.2& 0 \\
    \multicolumn{2}{c}{Random 90000} & 0.07 & 13 & 0 \\
    \bottomrule
    \end{tabular}
}
\vspace{-4mm}
\end{table}

\subsection{Ablation Study}\label{sec:evaluation_ablation_study}
This part evaluates \sysname's three core components. The enumerator, deduplicator, and ranker contribute to \sysname’s efficiency and effectiveness, with synergistic gains when combined.

\noindent \textbf{Enumerator.} As shown in Table \ref{tab:normalize_callbrator}, Standardized templates (OP1) reduce verification complexity from exponential to polynomial for redundant templates, cutting 3- and 4-node enumeration time by 44.86\% and 65.73\%, respectively, compared to no optimization (No OPT). This component alone reduces verification calls by 50.24\% and 68.01\% for 3- and 4-node templates, eliminating trivial structural variations (e.g., operator swaps) early in the process.

\begin{table*}[!htbp]
\centering
\caption{Higher ranking effective rules identified by \sysname, none of which were discovered by WeTune. 
Each $t_i$ signifies an input table, while each $a_i$ denotes a list of attributes, and each $p_i$ indicates a predicate. 
Multiple instances of the same symbol (i.e., $t_i$, $a_i$, $p_i$) illustrate equivalent constraints. Each $(t_i.a_j)$ denotes a constraint SubAttrs($a_j$, $a_{t_i}$). Additional types of constraints are detailed in the Extra Constraints column.}
\label{tab: example of rules} 
  \centering
  \small
  \setlength{\tabcolsep}{3pt}
  \renewcommand{\arraystretch}{1.2}
  \resizebox{1\textwidth}{!}{
  \begin{tabular}{| c | c | c | c |}
    \toprule
    No. & Source Plan Template & Destination Plan Template & Extra Constraints \\ 
    \midrule
     1 & $\mathrm{InSub}_{a_4}(t_0,\mathrm{Proj}_{a_1,s_1}(\mathrm{Sel}_{p_0,a_2}(\mathrm{InSub}_{a_1}(t_1,\mathrm{Proj}_{a_0,s_0}(t_2)))))$ & $\mathrm{InSub}_{a_4}(t_0,\mathrm{Proj}_{a_0,s_0}(\mathrm{Sel}_{p_0,a_2}(\mathrm{IJoin}_{a_1,a_0}(t_1,t_2))))$ & $\mathrm{ }$ \\   
    \hline
    2 & $\mathrm{InSub}_{a_2}(\mathrm{LJoin}_{a_0,a_0}(\mathrm{Proj}_{a_0,s_0}^{*}(\mathrm{InSub}_{a_0}(t_0,t_0)),\mathrm{Proj}_{a_2,s_1}^{*}(t_0)),t_0)$ & $\mathrm{IJoin}_{a_0,a_0}(\mathrm{Proj}_{a_0,s_0}^{*}(t_0),\mathrm{Proj}_{a_2,s_1}^{*}(t_0))$ & $\mathrm{NotNull}(t_0,a_2)$ \\
    \hline
    3 & $\mathrm{InSub}_{a_1}(\mathrm{LJoin}_{a_1,a_0}(t_0,\mathrm{Proj}_{a_1,s_0}^{*}(\mathrm{InSub}_{a_0}(t_0,t_0))),\mathrm{Proj}_{a_0,s_1}(t_0))$ & $\mathrm{IJoin}_{a_1,a_0}(t_0,\mathrm{Proj}_{a_1,s_0}^{*}(t_0))$ & $\mathrm{NotNull}(t_0,a_0)$ \\
    \hline
    4 & $\mathrm{LJoin}_{a_0,a_0}(\mathrm{Proj}_{a_1,s_0}^{*}(\mathrm{Sel}_{p_0,a_0}(t_0)),\mathrm{Proj}_{a_0,s_1}^{*}(\mathrm{Sel}_{p_0,a_0}(t_0)))$ & $\mathrm{IJoin}_{a_0,a_0}(\mathrm{Proj}_{a_1,s_0}^{*}(t_0),\mathrm{Proj}_{a_0,s_1}^{*}(\mathrm{Sel}_{p_0,a_0}(t_0)))$ & $\mathrm{NotNull}(t_0,a_0)$ \\
    \hline
    5 & $\mathrm{InSub}_{a_0}(\mathrm{IJoin}_{a_0,a_1}(\mathrm{Proj}_{a_1,s_0}^{*}(\mathrm{InSub}_{a_0}(t_0,t_0)),\mathrm{Proj}_{a_0,s_1}^{*}(t_0)),t_0)$ & $\mathrm{IJoin}_{a_0,a_1}(\mathrm{Proj}_{a_0,s_0}^{*}(t_0),\mathrm{Proj}_{a_1,s_1}^{*}(t_0))$ & $\mathrm{NotNull}(t_0,a_0)$ \\
    \hline
    6 & $\mathrm{IJoin}_{a_5,a_4}(\mathrm{Proj}_{a_1,s_1}(\mathrm{LJoin}_{a_1,a_1}(\mathrm{Proj}_{a_0,s_0}^{*}(t_0),t_0)),\mathrm{Proj}_{a_4,s_2}^{*}(t_2))$ & $\mathrm{IJoin}_{a_5,a_6}(\mathrm{Proj}_{a_1,s_1}(\mathrm{IJoin}_{a_1,a_1}(\mathrm{Proj}_{a_0,s_0}^{*}(t_0),t_0)),\mathrm{Proj}_{a_6,s_2}^{*}(t_2))$ & $\mathrm{ }$ \\
    \hline
    7  & $\mathrm{Sel_{p_0,t_0.a_4}(IJoin_{t_0.a_4,t_1.a_1}(t_0,Proj_{t_1.a_1}(IJoin_{t_1.a_1,t_2.a_0}(t_1,Dedup(Proj_{t_2.a_0}(t_2)))}$    
    & $\mathrm{IJoin_{t_0.a_4,t_1.a_1}(t_0,Proj_{t_1.a_1}(Sel_{p_0,t_1.a_1}(InSub_{t_1.a_1}(t_1,Proj_{t_2.a_0}(t_2)))))}$     
    &$\mathrm{ }$\\ 
    \hline
    8 & $\mathrm{Sel}_{p_0,a_0}(\mathrm{InSub}_{a_0}(\mathrm{IJoin}_{a_0,a_0}(\mathrm{Proj}_{a_0,s_0}(\mathrm{Sel}_{p_0,a_0}(t_0)),t_0),t_0))$ & $\mathrm{LJoin}_{a_0,a_0}(\mathrm{Proj}_{a_0,s_0}^{*}(\mathrm{Sel}_{p_0,a_0}(t_0)),t_0)$ & $\mathrm{NotNull}(t_0,a_0);\mathrm{Unique}(t_0,a_0)$ \\
    \hline
    9 & $\mathrm{Proj}_{a_5,s_1}^{*}(\mathrm{InSub}_{a_0}(\mathrm{IJoin}_{a_0,a_1}(t_0,t_1),\mathrm{Proj}_{a_2,s_0}(\mathrm{Sel}_{p_0,a_2}(t_2))))$ & $\mathrm{Proj}_{a_5,s_1}(\mathrm{InSub}_{a_0}(t_0,\mathrm{Proj}_{a_2,s_0}(\mathrm{Sel}_{p_0,a_1}(\mathrm{LJoin}_{a_1,a_2}(t_1,t_2)))))$ & $\mathrm{Unique}(t_0,a_5)$ \\
    \hline
    10 & $\mathrm{IJoin}_{a_0,a_1}(\mathrm{Proj}_{a_1,s_0}^{*}(\mathrm{InSub}_{a_0}(t_0,t_0)),\mathrm{Proj}_{a_0,s_1}(\mathrm{Sel}_{p_0,a_1}(t_0)))$ & $\mathrm{IJoin}_{a_0,a_1}(\mathrm{Proj}_{a_0,s_0}^{*}(\mathrm{Sel}_{p_0,a_0}(t_0)),\mathrm{Proj}_{a_1,s_1}(\mathrm{Sel}_{p_0,a_0}(t_0)))$ & $\mathrm{NotNull}(t_0,a_0)$ \\ 
    \hline
    11 &
    \makecell[c]{
      $\mathrm{InSubFilter}_{a_5}(\mathrm{InnerJoin}_{a_1,a_2}(\mathrm{Input}_{t_0},\mathrm{Proj}^*_{a_0,s_0}(\mathrm{Input}_{t_1})),\ \mathrm{Proj}_{a_5,s_1}($ \\
      $\mathrm{LeftJoin}_{a_1,a_2}(\mathrm{Input}_{t_0},\mathrm{Input}_{t_1})))$
    } &
    \makecell[c]{
      $\mathrm{LeftJoin}_{a_1,a_2}(\mathrm{Input}_{t_0},\mathrm{Proj}^*_{a_0,s_0}(\mathrm{Input}_{t_1}))$ 
    } &
    \makecell[c]{
      $\mathrm{NotNull}(t_0,a_1);\mathrm{NotNull}(t_0,a_5);$ \\
      $\mathrm{Reference}(t_0,a_1,t_1,a_2)$
    } \\

    \hline

    12 & $\mathrm{IJoin}_{a_2,a_0}(\mathrm{Proj}_{a_2,s_0}(\mathrm{IJoin}_{a_0,a_1}(t_0,t_1)),\mathrm{Proj}_{a_0,s_1}(\mathrm{Sel}_{p_0,a_0}(t_0)))$ 
    & $\mathrm{IJoin}_{a_5,a_0}(\mathrm{Proj}_{a_0,s_0}(t_0),\mathrm{Proj}_{a_1,s_1}(\mathrm{Sel}_{p_0,a_0}(\mathrm{IJoin}_{a_1,a_0}(t_1,t_0))))$ 
    & $\mathrm{Unique}(t_0,a_0)$ \\ 
    
    \hline
    
    13 & $\mathrm{InSub}_{a_0}(\mathrm{LJoin}_{a_0,a_1}(\mathrm{Proj}_{a_1,s_0}(\mathrm{Sel}_{p_0,a_0}(t_0)),\mathrm{Proj}^*_{a_0,s_1}(t_0)),t_0)$ 
    & $\mathrm{IJoin}_{a_0,a_1}(\mathrm{Proj}_{a_1,s_0}(\mathrm{InSub}_{a_0}(t_0,t_0)),\mathrm{Proj}^*_{a_1,s_1}(\mathrm{Sel}_{p_0,a_0}(t_0)))$ 
    & $\mathrm{Unique}(t_0,a_1)$ \\ 
    
    \hline
    
    14 & $\mathrm{LJoin}_{a_0,a_0}(\mathrm{Proj}_{a_0,s_0}(\mathrm{Sel}_{p_0,a_0}(t_0)),\mathrm{Proj}^*_{a_3,s_1}(\mathrm{InSub}_{a_0}(t_0,t_0)))$ 
    & $\mathrm{IJoin}_{a_0,a_0}(\mathrm{Proj}_{a_0,s_0}(t_0),\mathrm{Proj}^*_{a_3,s_1}(\mathrm{Sel}_{p_0,a_3}(t_0)))$ 
    & $\mathrm{NotNull}(t_0,a_0);\mathrm{Unique}(t_0,a_0)$ \\ 
    
    \hline
    
    15 & $\mathrm{LJoin}_{a_0,a_1}(\mathrm{Proj}_{a_1,s_0}(\mathrm{Sel}_{p_0,a_0}(t_0)),\mathrm{Proj}^*_{a_0,s_1}(\mathrm{IJoin}_{a_1,a_3}(t_0,t_2)))$ 
    & $\mathrm{IJoin}_{a_0,a_1}(\mathrm{Proj}_{a_0,s_0}(t_0),\mathrm{Proj}_{a_0,s_1}(\mathrm{Sel}_{p_0,a_1}(t_0)))$ 
    & $\mathrm{Ref}(t_0,a_1,t_2,a_3);\mathrm{Unique}(t_0,a_0)$ \\ 
    
    \hline
    
    16 & $\mathrm{InSub}_{a_3}(\mathrm{LJoin}_{a_3,a_0}(\mathrm{Proj}_{a_0,s_0}(\mathrm{LJoin}_{a_0,a_1}(t_0,t_1)),\mathrm{Proj}^*_{a_3,s_1}(t_0)),t_3)$ 
    & $\mathrm{IJoin}_{a_4,a_0}(\mathrm{Proj}_{a_0,s_0}(\mathrm{LJoin}_{a_0,a_1}(t_0,t_1)),\mathrm{Proj}_{a_3,s_1}(\mathrm{InSub}_{a_3}(t_0,t_3)))$ 
    & $\mathrm{Ref}(t_1,a_1,t_0,a_0);\mathrm{Unique}(t_0,a_0)$ \\ 
    
    \hline
    
    17 & $\mathrm{InSub}_{a_2}(\mathrm{LJoin}_{a_2,a_0}(\mathrm{Proj}_{a_0,s_0}(\mathrm{Sel}_{p_0,a_0}(t_0)),\mathrm{Proj}^*_{a_2,s_1}(t_0)),t_0)$ 
    & $\mathrm{LJoin}_{a_2,a_0}(\mathrm{Proj}^*_{a_0,s_0}(\mathrm{Sel}_{p_0,a_0}(t_0)),\mathrm{Proj}^*_{a_2,s_1}(t_0))$ 
    & $\mathrm{Unique}(t_0,a_0)$ \\ 
    
    \hline
    
    18 & $\mathrm{LJoin}_{a_0,a_2}(\mathrm{Proj}_{a_2,s_0}(\mathrm{IJoin}_{a_0,a_1}(t_0,t_1)),\mathrm{Proj}^*_{a_0,s_1}(\mathrm{InSub}_{a_2}(t_0,t_0)))$ 
    & $\mathrm{IJoin}_{a_0,a_6}(\mathrm{Proj}_{a_0,s_0}(t_0),\mathrm{Proj}_{a_1,s_1}(\mathrm{InSub}_{a_1}(\mathrm{LJoin}_{a_1,a_2}(t_1,t_0),t_0)))$ 
    & $\mathrm{Unique}(t_0,a_2)$ \\ 
    
    \hline
    
    19 & $\mathrm{Sel}_{p_0,a_1}(\mathrm{LJoin}_{a_0,a_1}(\mathrm{Proj}_{a_1,s_0}(\mathrm{Sel}_{p_0,a_0}(t_0)),\mathrm{Proj}^*_{a_0,s_1}(t_0)))$ 
    & $\mathrm{IJoin}_{a_0,a_1}(\mathrm{Proj}^*_{a_0,s_0}(\mathrm{Sel}_{p_0,a_0}(t_0)),\mathrm{Proj}^*_{a_0,s_1}(t_0))$ 
    & $\mathrm{Unique}(t_0,a_1)$ \\ 
    
    \hline
    
    20 & $\mathrm{InSub}_{a_0}(\mathrm{IJoin}_{a_1,a_0}(\mathrm{Proj}_{a_0,s_0}(t_0),\mathrm{Proj}_{a_1,s_1}(t_0)),\mathrm{Proj}_{a_1,s_2}(t_0))$ 
    & $\mathrm{IJoin}_{a_0,a_0}(\mathrm{Proj}_{a_1,s_0}(t_0),\mathrm{Proj}_{a_0,s_1}(t_0))$ 
    & $\mathrm{Unique}(t_0,a_0)$ \\

    \bottomrule
  \end{tabular}
  }
\end{table*}

\noindent  \textbf{Deduplicator.} As shown in Table \ref{tab:normalize_callbrator}, RTP (OPT2) prunes redundant rules during enumeration, further reducing checker calls by 3.82–4.02\% for 2–4 node templates and cutting 4-node deduplication time from 33 minutes (\sysname without RTP) to <10 minutes (\sysname), as shown in Table \ref{tab:normalize_stat}. It ensures only redundant rules are removed, preserving optimization capacity while minimizing overhead. Nodes exceeding 5 are excluded, as the baseline (without OPT1 and OPT2) fails to complete enumeration for such cases.

Table \ref{tab:normalize_rules} summarizes the overhead of enumeration and deduplication. Generating a standardized rule base for templates with up to 6 nodes incurs minimal overhead due to its linear complexity (traversing each template once, proportional to template count). 
Even as the maximum node count increases (raising average nodes per template), the enumeration and de-redundancy overhead remain insignificant  (<0.05\% for 4- or more-node enumeration). 

\begin{figure}[t]
  \centering
  \includegraphics{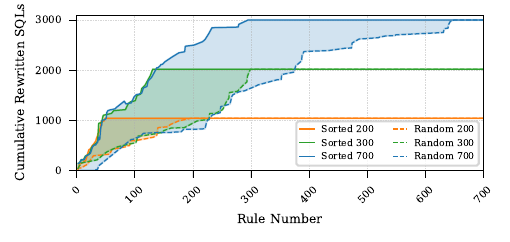}
  \caption{
  Evaluation of \sysname's ranker in terms of cumulative rewritten SQLs by adding rules.
  }\label{fig:rules sorted}
\end{figure}

\noindent \textbf{Ranker.} 
The ranker filters trivial rules, ensuring top-ranked rules deliver meaningful performance gains. 
Validation via the SQL Server optimizer shows that effective optimization rules are concentrated in top-ranked sets. 
As shown in Table \ref{tab:rule_effectiveness}, combinations of 50–1,000 top-ranked rules (paired with 0–200 low-ranked rules) achieve a 100\% trigger rate, with top-rule usage rates (proportion of top-ranked rules applied by the optimizer) of 11.4–52.0\% and no low-ranked rules applied.
Random experiments (average of 5 repeats) reinforce that, as shown in Figure \ref{fig:rules sorted}, in tests with 700 mixed rules, sorted subsets (700, 300, 200) completed rewrites for 3,000 queries by the 295th, 131st, and 49th rules, respectively, far outperforming random subsets (requiring 650th, 299th, 197th rules). This stems from prioritizing effective rules early.
Notable valid previous-not-found rules (Table~\ref{tab: example of rules}) are consistently top-ranked and frequently applied, significantly boosting rewriting efficiency for complex queries (e.g., Table~\ref{tab: example of query}). Noably, the ranker’s overhead (<1 minute) is negligible for overall performance.

\subsection{End-to-End Performance Impact of Rewrite Rules}\label{sec:End-to-End}

\begin{figure}[t]
  \centering
  \includegraphics{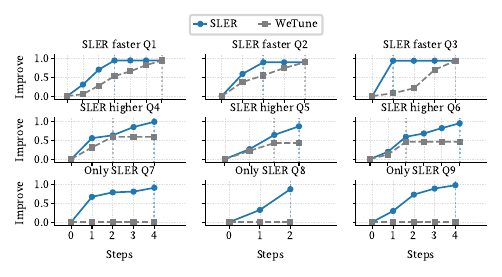}
  \caption{End-to-end latency improvement comparison.}
  \label{fig:endtoend}
\end{figure}

To further evaluate \sysname against no rewrite rules and WeTune, based on our evaluation, 9 representative queries are selected to demonstrate superior results as shown in Figure \ref{fig:endtoend}, which are categorized into three scenarios.

\textbf{Scenario A: Faster Convergence (Q1--Q3).} In these cases, both systems reach the same optimal plan. However, \sysname achieves this in significantly fewer steps. This is because \sysname's large-node rules act as ``composite transformations'' that encapsulate multiple small-node rules. While WeTune must apply several atomic rules sequentially, \sysname can bridge the gap in a single step, avoiding complex intermediate states.

    
\textbf{Scenario B: Higher Optimization (Q4--Q6).} \sysname achieves a higher improvement ratio than WeTune. as SLER proactively enumerates and verifies complex patterns that exceed the 4-node threshold. WeTune stalls at local optima because its small-rule library cannot perceive the deeper structural simplifications possible with larger templates.
    
\textbf{Scenario C: Unique Optimization Capability (Q7--Q9).} In these challenging cases, WeTune fails to provide any optimization (0\% improvement), whereas \sysname identifies significant simplification opportunities. This highlights the superior coverage of our rule library. Many intricate query patterns require specific multi-operator templates (e.g., large-node composite structures such as multi-level join-filter-join redundancies involving 5--9 nodes) that only \sysname's active discovery framework can successfully enumerate and verify.


Note that we also evaluated \sysname on MySQL; the optimization patterns and performance gains were consistent with those reported here.

\noindent \textbf{Performance Improvement for Top Rules.} 
Evaluation results confirm top-ranked rules from \sysname drive exceptional performance gains: the top 500 effective rules achieve a 98.6\% average improvement, vastly outperforming the bottom 50\% (11.9\% average). This stark disparity validates the ranking mechanism’s strong ability to prioritize high-value, latency-reducing rules from the large enumeration space, demonstrating the ranker’s high real-world efficacy in identifying impactful transformations that deliver positive optimization for evaluated workloads.

\end{sloppypar}

%% file: Conclusion.tex

This paper proposes \sysname, a rewrite framework via standardized templates, RTP deduplication, and LambdaMART ranking. It overcomes the ``small-to-large'' hypothesis, and cuts verification overhead by pruning redundancy, while prioritizing effective rules, reducing redundant template enumeration complexity from exponential to polynomial. Experimental results demonstrate that after training on over open-source and commercial queries, \sysname outperforms SOTA methods in both efficiency and rule effectiveness. 